\documentclass{aa}
\usepackage{graphicx}
\usepackage{txfonts}
\usepackage{hyperref}

\def\aap{A\&A}
\def\apjs{ApJS}

\newcommand\PEGASE{\textrm{\textsc{P\'egase}}}

\newcommand\df{\mathrm{d}}

\newcommand\mult{\mathclose{}\,\mathopen{}}

\newcommand\neper{\mathrm{e}}

\newcommand\micron{\mathrm{\muup m}}
\newcommand\Gyr{\mathrm{Gyr}}
\newcommand\Myr{\mathrm{Myr}}
\newcommand\parsec{\mathrm{pc}}

\newcommand\system{\mathrm{sys}}
\newcommand\infall{\mathrm{in}}
\newcommand\outflow{\mathrm{out}}
\newcommand\ISM{\textsc{ism}}
\newcommand\total{\mathrm{tot}}
\newcommand\galaxy{\mathrm{gal}}
\newcommand\stel{\star}
\newcommand\dust{\mathrm{d}}
\newcommand\SSP{\textsc{ssp}}

\newcommand\extin{\mathrm{ext}}
\newcommand\absor{\mathrm{abs}}
\newcommand\scat{\mathrm{sca}}

\newcommand\singleStar{\mathrm{star}}
\newcommand\Teff{T_{\mathrm{eff}}}
\newcommand\Lbol{\mathcal{L}}

\newcommand\compos{\chi}
\newcommand\composInit{\compos_0^{}}
\newcommand\composSurf{\compos_{\mathrm{surf}}^{}}
\newcommand\carbon{\mathrm{carb}}
\newcommand\silic{\mathrm{sil}}

\newcommand\disk{\mathrm{disk}}
\newcommand\bulge{\mathrm{bulge}}
\newcommand\nSersic{n_{\mathrm{S}}}
\newcommand\spiral{\mathrm{spir}}

\newcommand\spher{\mathrm{sph}}
\newcommand\nFroehlich{n_{\mathrm{F}}}
\newcommand\Rcore{R_{\mathrm{c}}}
\newcommand\Rtrunc{R_{\mathrm{t}}}

\newcommand\LymCont{LC}
\newcommand\cavity{\mathrm{c}}
\newcommand\rSz{r_{\mathrm{S},\,0}}
\newcommand\rSd{r_{\mathrm{S}}}
\newcommand\rCav{r_{\cavity}}
\newcommand\gammaSd{\gamma_{\mathrm{S}}}
\newcommand\gammaCav{\gamma_{\cavity}}
\newcommand\epsilonSd{\varepsilon_{\mathrm{S}}}
\newcommand\epsilonCav{\varepsilon_{\cavity}}
\newcommand\NLC{N}
\newcommand\nH{n_{\mathrm{H}}}
\newcommand\tauSd{\zeta}
\newcommand\HII{\ion{H}{ii}}
\newcommand\HI{\ion{H}{i}}
\newcommand\tauH{\tau_{\mathrm{H}}}
\newcommand\kappaH{\kappa_{\mathrm{H}}}
\newcommand\taud{\tau_\dust}
\newcommand\alphaB{\alpha_{\mathrm{B}}}
\newcommand\nElec{n_{\mathrm{e}}}
\newcommand\model{\mathrm{mod}}
\newcommand\cluster{\textsc{sc}}
\newcommand\birthCloud{\textsc{bc}}
\newcommand\DISM{\textsc{dism}}
\newcommand\SFC{\textsc{sfc}}

\newcommand\Latt{L}
\newcommand\meanLatt{\bar L}
\newcommand\unatt{0}
\newcommand\Lunatt{L^{\unatt}}

\newcommand\BoverT{\Gamma}
\newcommand\OtoSil{\Xi}
\begin{document}
\title{\PEGASE.3: A code for modeling the \mbox{UV-to-IR/submm} spectral\\
  and chemical evolution of galaxies with dust}
\titlerunning {\PEGASE.3: spectrochemical evolution of galaxies with dust}
\authorrunning{M.~Fioc \&  B.~Rocca-Volmerange}
\author{Michel Fioc
  \inst{1}
  \and
  Brigitte Rocca-Volmerange
  \inst{1,2}
}
\institute{Sorbonne Universit\'e, CNRS, UMR~7095, Institut d'astrophysique de Paris, 98 bis, b\textsuperscript{d} Arago, 75014 Paris, France\\
  \email{Michel.Fioc@iap.fr},
  \and
  Universit\'e Paris-Sud, 91405 Orsay, France\\
  \email{Brigitte.Rocca@iap.fr}
}
\date{Received         ; accepted         }
\abstract{%
  A code computing consistently the evolution of stars, gas and dust,
  as well as the energy they radiate, is required to derive reliably
  the history of galaxies by fitting synthetic spectral energy
  distributions (SEDs) to multiwavelength observations.
  The new code \PEGASE.3 described in this paper extends to
  the far-infrared\slash submillimeter the
  ultraviolet-to-\mbox{near-infrared} modeling provided by previous versions
  of \PEGASE.
  It first computes the properties of single stellar populations
  at various metallicities.
  It then follows the evolution of the stellar light of a galaxy
  and the abundances of the main metals in the interstellar medium (ISM),
  assuming some scenario of mass assembly and star formation.
  It simultaneously calculates the masses of the various grain families, the
  optical depth of the galaxy and the attenuation of the SED through
  the diffuse ISM in spiral and spheroidal galaxies, using grids of
  radiative transfer precomputed with Monte Carlo simulations taking
  scattering into account.
  The code determines the mean radiation field and the temperature probability
  distribution of stochastically heated individual grains.
  It then sums up their spectra to yield the overall emission by dust in
  the diffuse ISM.
  The nebular emission of the galaxy is also computed, and
  a simple modeling of the effects of dust on the SED of star-forming regions
  is implemented.

  The main outputs are ultraviolet-to-submillimeter SEDs of galaxies
  from their birth up to $20\mult\Gyr$, colors, masses of galactic components,
  ISM abundances of metallic elements and dust species,
  supernova rates.
  The temperatures and spectra of individual grains are also available.
  The paper discusses several of these outputs for a scenario representative
  of Milky Way-like spirals.

  \PEGASE.3 is fully documented and its Fortran~95 source files are public.
  The code should be especially useful for cosmological simulations
  and to interpret future mid-~and far-infrared data, whether obtained by
  JWST, LSST, Euclid or e-ELT.
}
\maketitle
\section{Introduction}
Multiwavelength observations of galaxies
from the far-ultraviolet to the far-infrared and submillimeter
\mbox{--~}the spectral domain in which ordinary galaxies emit almost all their light~--
should allow reconstruction of
their history and that of their stars, gas and dust.
Many deep surveys have already been carried out to derive the evolution
on cosmological timescales of the stellar mass for a large number of objects
of various types.
In particular, determining the history of mass assembly of galaxies might help
to assess the respective contributions of hierarchical merging
and early dissipative collapse to the formation of galaxies,
and, ultimately, to settle the long-standing astrophysical debate over which was the dominant process.

To this purpose, several teams have built codes to synthesize panchromatic
spectral energy distributions (SEDs), fit them to observations of galaxies,
whether individual or in surveys, and derive their
ages, masses, star formation rates, chemical compositions, and opacities
(e.g., the code \textsc{Beagle} by \citealp{Chevallard+Charlot2016}, to name
a recent one).
A first difficulty is to jointly follow the evolution of gas and dust in
the interstellar medium (ISM) and that of the stellar populations
that form from the latter and enrich it.
Another difficulty is to model properly the attenuation and scattering by
dust grains of the light emitted by stars in the ultraviolet (UV) and the
optical, as well as the re-emission at longer wavelengths of the energy
these grains absorbed.
As discovered by \emph{IRAS} and confirmed by \emph{ISO}, \emph{Spitzer} and
\emph{Herschel}, this process dominates at all wavelengths from the
mid-~to the far-infrared (IR).
It is therefore essential to model it consistently to put reliable constraints
on the evolution of galaxies using synthetic SEDs.
This is the main objective of the code \PEGASE.3 described in this paper.

A large variety of spectral synthesis codes already exist
(for review papers, see \citealp{Walcher2011} and \citealp{Conroy_ARAA}).
Some, such as \textsc{Starburst99} \citep{Leitherer1999, Leitherer2014}
and codes based on it \citep[e.g.,][]{Steidel2016}, focus on the modeling
of young stellar populations ($\mathord{\la}\,100\mult\Myr$) and their
nebular environment;
they are typically applied to the analysis of the continuum
and of emission lines at UV--optical wavelengths.
The code of \citet{Dopita2005b}, which couples \textsc{Starburst99}
and \textsc{Mappings}, also computes the infrared emission produced
by the dust cloud surrounding a star cluster.
Although these codes are very appropriate to study current or recent starbursts,
they are not suitable for evolved galaxies.

The first code following the evolution of the stellar light emitted
by a galaxy on cosmological timescales was built by \citet{Tinsley1972}.
It computed the mass and overall metallicity of the interstellar medium
as a function of time, and implemented laws relating the star formation rate
to the gas content.
This code produced however only optical colors.

The spectral evolution code of \citet{Bruzual1983b} was a significant
improvement, not only with regard to the photometric evolution codes
of \citet{Tinsley1972} and \citet{RV81}, but also when compared to
then-existing spectral population synthesis codes
\citep[e.g.,][]{Alloin+1971}, which simply fitted the SEDs of galaxies
with linear combinations of observed SEDs of stars or star clusters.

The extension to the near-infrared was a difficult challenge,
because it required to follow the rapid stellar phases
(the late asymptotic giant branch, mainly) dominating these wavelengths.
This problem was solved by \citet{CB1991} through the method of isochrones,
also used since the first version of code~\PEGASE\ \citep{Pegase.1}.
Another solution, based on the fuel consumption theorem
\citep[see][for instance]{RenziniBuzzoni},
was implemented by \citet{Maraston2005}.

Many other developments occurred meanwhile.
To cite a few, \citet{GRV1987} modeled the metallicity-dependent
attenuation of a galaxy SED by dust grains distributed in a slab;
\citet{Worthey} considered the effect of non-solar abundance ratios of
metals on the spectrum of a single stellar population;
\citeauthor{Pegase.2} (\citeyear{Pegase.2}; \PEGASE.2) computed consistently
the metallicity and the spectral evolution, from the UV to the \mbox{near-IR},
using metallicity-dependent stellar spectra, evolutionary tracks and yields;
\citet{Devriendt1999} developed the code \textsc{Stardust} and used it in
\textsc{Galics} \citep{Hatton+2003,Cousin+2015} to run semi-analytic
simulations of galaxy formation;
\citet{Boissier} modeled in detail the radial dependency of the spectrochemical
evolution of spiral galaxies;
\citet{Eldridge, Eldridge+2017} explored the effects of binaries.

To extend spectral evolution codes to longer wavelengths, it was necessary
to model properly dust grains and their effects in attenuation and
emission.
Fits of the Milky Way's extinction curve by \citet{Mathis1977} initially
suggested that dust was a mixture of silicate and graphite grains
with sizes larger than $0.005\mult \micron$, but the analysis of the
\mbox{mid-IR} features observed in reflection nebulae
led \citet{Sellgren1984} to propose that these were caused by much
smaller, stochastically heated particles.
\citet{LegerPuget1984} subsequently identified these particles
as polycyclic aromatic hydrocarbons (PAHs);
these molecules are now commonly included in models of dust composition
and grain size distribution \citep[e.g.,][]{Zubko2004,Weingartner2001}.
The evolution of the two main families of grains, namely carbonaceous and
silicate ones, was modeled by \citet{Dwek1998}, taking into account their
formation, destruction and accretion on preexisting grains.

Several galaxy evolution models compute the attenuation by dust of the
UV--optical light using global effective attenuation curves such as the ones
proposed by \citeauthor{Calzetti1994} (\citeyear{Calzetti1994}; recently
updated by \citealp{Battisti2016}), \citet{CharlotFall2000} or
\citet{Conroy2010}.
Effective attenuation curves were also used by \citet{LoFaro2017}
to fit the UV--IR SED of galaxies with \textsc{Cigale} \citep{Buat2014}, a code
enforcing the energy balance between dust absorption and dust emission.

The energy balance argument is also invoked in the codes \textsc{Stardust}
\citep{Devriendt1999} and \textsc{Magphys} \citep{DaCunha2008} to estimate
the integrated emission by dust;
this emission is then apportioned among predefined SEDs corresponding to
PAHs, small and big graphite and silicate grains with temperatures fixed
beforehand.
These codes do not, however, compute the temperature distribution of individual
grains stochastically heated by the radiation field.
Neither do they follow the evolution of the dust content in the ISM.

True radiative transfer codes are required to determine the attenuation
and emission by dust more reliably.
Several techniques have been developed to this purpose:
\textsc{Grasil} \citep{Granato2000,Lacey2008}, for instance, is based
on Monte Carlo simulations;
on the other hand, the recent modeling by \citet{Cassara2015}
uses the ray-tracing method.

All codes have to make compromises between efficiency, completeness,
accuracy and consistency.
Some of the codes listed above are more physical than \PEGASE.3
in some respects, but are so numerically intensive that they are
restricted in practice to the study of single objects or specific
categories of galaxies.
Others are more phenomenological and, while they may be very handy
to analyze large surveys,
give little insight on the evolution of galaxies because the various
components, in particular stars and dust, are modeled separately;
they have therefore little predicting power.

\PEGASE.3 aims to reconcile efficiency, consistency and realistic modeling
in a general-purpose code, appropriate for studies of
the chemical and \mbox{far-UV}-to-submillimeter (submm) spectral evolution
of Hubble sequence galaxies from their formation up to now.
The variety of the inputs used to compute galaxy models,
the wealth of its outputs and its modular structure should make it
particularly suited to cosmological simulations.
Its source files, written in Fortran~95, are moreover entirely public%
\footnote{Available at
  \ifdefined\href
  \href{http://www.iap.fr/users/fioc/Pegase/Pegase.3/}{\texttt{www.iap.fr\slash users\slash fioc\slash Pegase\slash Pegase.3/}}%
  \else
  \texttt{www.iap.fr\slash users\slash fioc\slash Pegase\slash Pegase.3/}%
  \fi\
  and
  \ifdefined\href
  \href{http://www.iap.fr/pegase/}{\texttt{www.iap.fr\slash pegase/}}%
  \else
  \texttt{www.iap.fr\slash pegase/}%
  \fi\,.}
and may be freely adapted by the user.
For technical details, see the documentation \citep{doc_Pegase.3}.

Section~\ref{sec:spectral_chemical_model} of this paper
first describes the system considered by \PEGASE.3,
and how the code consistently computes the spectral evolution
of a galaxy's stellar component and the chemical evolution of its interstellar
medium (ISM) for a given scenario of mass assembly and star formation.
We detail in particular the inputs used to model single stellar populations.
Section~\ref{sec:grains} is devoted to the characteristics of grains
\mbox{--~}their optical properties and size distribution~--
and to their evolution in the ISM.
Section~\ref{sec:diffuse_attenuation} focuses on the radiative transfer
of the stellar light through the dusty diffuse ISM in spiral and spheroidal
galaxies.
The computation of dust emission, taking into account the stochastic heating of
grains by the radiation field, is treated in Sect.~\ref{sec:diffuse_emission}.
The modeling of nebular emission and of star-forming clouds is dealt with
in Sect.~\ref{sec:cloud}.
In Sect.~\ref{sec:outputs}, we highlight some outputs of \PEGASE.3, taking
as an example a scenario representative of Milky Way-like galaxies.
Finally, Sect.~\ref{sec:conclusion}
concludes on the strengths and limitations of the code,
its existing and potential applications, and lists some intended improvements.
\section{Emission of stellar populations and chemical evolution%
  \label{sec:spectral_chemical_model}}
\subsection{The system%
  \label{sec:system}}
The system considered in the code is formed of the galaxy proper, of
reservoirs furnishing the gas falling onto the galaxy (infall),
and of interstellar matter ejected by the galaxy into the intergalactic
medium (outflow):
so, the galaxy is modeled as an open box, but the total mass of the system,
$M_\system$, is constant.
The evolution of the galaxy from its initial state is determined by a scenario
providing, among others, the parameters from which the star formation, infall
and outflow rates are computed as a function of time.
All the parameters defining a scenario are described in detail
in the code's documentation \citep{doc_Pegase.3};
they are organized in trees for convenience.

The code follows the evolution of the interstellar medium (ISM; both gas and
dust), of the stars it contains and of the mass locked in
compact stellar remnants (black holes and neutron stars).
Two regions are distinguished in the ISM: the diffuse medium and star-forming
clouds (see \citealp{Silva1998} and \citealp{CharlotFall2000} for other
instances of this distinction).
\subsection{Modeling of single stellar populations}
\subsubsection{Overview}
The basic unit of a galaxy evolution model is a single stellar population
(SSP), that is, a set of stars, with the same initial chemical composition,
created by an instantaneous star-forming event.
The monochromatic luminosity (or ``spectral flux'', in radiometric terminology;
denoted by $L_\lambda$ hereafter) per unit wavelength of an SSP with an initial
chemical composition
$\{\composInit\} \coloneqq \{\composInit(\mathrm{H}), \composInit(\mathrm{He}), \ldots\}$
is, per unit initial mass of the SSP,
\begin{equation}
  \label{eq:integr_isoch}
  L_\lambda^\SSP(t, \{\composInit\}) =
  \int L_\lambda^\singleStar(m, t, \{\composInit\})\mult \phi(m)\mult \df(\ln m)
\end{equation}
at age~$t$ and wavelength~$\lambda$, where
$L_\lambda^\singleStar(m, t, \{\composInit\})$ is the monochromatic luminosity
at this age of a star with an initial mass~$m$ and initial
composition~$\{\composInit\}$, and $\phi$ is the initial mass function
(IMF): $\phi(m)\mult \df(\ln m)$ is the number of stars, per unit initial mass
of the SSP, born with a mass in the interval $[m, m+\df m\mathclose{[}$;
this function is normalized, that is, $\int m\mult \phi(m)\mult \df(\ln m) = 1$.
A large number of IMFs are available in the code.

The luminosity $L_\lambda^\singleStar$ is computed as
\begin{equation}
  L_\lambda^\singleStar(m, t, \{\composInit\})
  = \Lbol_\singleStar(m, t, \{\composInit\})\mult
  \ell_\lambda^\singleStar(\{\composSurf\}, \Teff, g)\,,
\end{equation}
where $\Lbol_\singleStar$ is the bolometric luminosity (or ``radiant flux'',
in radiometric terminology;
all integrated luminosities are denoted by $\Lbol$ hereafter) of the star,
that is, the amplitude of the stellar spectrum, and the function
$\lambda \mapsto \ell_\lambda^\singleStar$ is the shape of this spectrum.
This shape depends mainly%
\footnote{The effects of rotation and stellar winds on stellar spectra are neglected.}
on three quantities:
the surface composition $\{\composSurf\}$ of the star,
its effective temperature $\Teff$
and its surface gravity $g$.
Because of limitations in the input data, we use a single number,
the initial metallicity $Z$ of a star
\mbox{--~}its mass fraction of metals at birth~\mbox{--,}
as a substitute for both the initial and surface sets of abundances,
$\{\composInit\}$ and $\{\composSurf\}$.
The quantities $\Lbol_\singleStar$, $\Teff$ and $g$ are given by
stellar evolutionary tracks as a function of~$m$, $t$ and~$Z$.
The shape is interpolated as a function of $Z$, $\Teff$ and $g$
from the elements of a metallicity-dependent library of stellar spectra.
To compute $L_\lambda^\SSP$ from Eq.~\eqref{eq:integr_isoch}, we derive from
the evolutionary tracks the isochrone of an SSP at age~$t$,
that is, the locus of all the stars in the $(\Lbol_\singleStar, \Teff, g)$
($\mathord{\approx}\,$Hertzsprung \& Russell; HR) diagram.
\subsubsection{Stellar evolutionary tracks}
Stellar evolutionary tracks provide the evolution of the bolometric
luminosity, effective temperature and surface gravity as a function
of age for a range of initial masses.
They should ideally be available for all the initial compositions
(or initial metallicities, at least) occurring
during the evolution of a galaxy, cover all the masses in the IMF
and all the evolutionary phases of a star.

The default set is based on the classical ``Padova'' tracks
\citep{Padova0.02, Padova0.0004-0.05,
  Padova0.004-0.008, Padova0.1, Padova0.0001}%
\footnote{More-recent sets of tracks exist, for instance \textsc{Parsec} by
  the same team \citep{PARSEC}, but the classical set is still the
  one with the largest extent in metallicity.},
from $Z = 0.0001$ up to $Z = 0.1$.
For stars undergoing the helium flash, the mass loss along the
red (or ``first'') giant branch (RGB)
and early asymptotic giant branch (AGB) phases is modeled with the law of
\citet{Reimers1975} and multiplied by an efficiency~$\eta$ \citep{Renzini1981};
we take $\eta = 0.4$, as recommended by the authors of the Padova tracks.
Pseudo-tracks are then computed for the thermally pulsing AGB phase,
using the equations proposed by \citet{Groenewegen1993} with $\eta = 4$
\citep{vdH1997}.
Hydrogen-burning post-AGB and CO white dwarf tracks are derived from
\citet{Bloecker1995}, \citet{Schoenberner1983}, \citet{Koester1986} and
\citet{Paczynski1971}.
For low-mass stars becoming helium white dwarfs, the \citet{Althaus1997}
models are used, while the nearly unevolving positions of low-mass stars
($m < 0.6\mult M_\odot$) in the HR diagram come from \citet{Chabrier1997}.
\subsubsection{Stellar yields%
  \label{sec:SSPs_yields}}
Stars eject matter in the interstellar medium (ISM) through stellar winds
and when they explode as supernovae.
When the code computes the chemical evolution of a galaxy,
it is assumed that the ejection of matter by a star happens only at the end
of its life:
this approximation is justified by the short life of high-mass stars,
compared to the age of a galaxy,
and by the late onset of intense winds in low-mass stars.
The recycling of matter in galaxies modeled by \PEGASE\ is therefore
not instantaneous.

For low-mass stars ($m/M_\odot \leqslant 5$ or~$6$ in Padova tracks),
stellar winds occur mainly during the RGB and AGB phases;
the final remnant is a white dwarf.
We used the yields in tables~A7 to~A12
of \citet{Marigo2001},
given for a mixing-length parameter $\alpha = 1.68$ at
$Z \in \{0.004, 0.008, 0.019\}$, and extrapolated
them at the metallicities of the Padova tracks.

Higher-mass stars undergo strong stellar winds during their whole life and
usually end as core-collapse supernovae (i.e., of type~II, Ib or Ic);
the final remnant is a neutron star or a stellar black hole,
depending on the progenitor's mass.
For these stars, the code's default yields are from \citet{Portinari1998}.
The choice of the supernova yields, with no winds, from model~B of
\citet{Woosley1995} is also possible, but we had to extrapolate them above
an initial mass of $40\mult M_\odot$ and below $11\mult M_\odot$.
Files containing all the computed yields are available with the code.

Although more-recent yields have been published
(see the review by \citealp{Nomoto+2013}, in particular for high-mass stars,
and, for low-mass stars, \citealp{Karakas2010}),
the yields of \citet{Marigo2001} and \citet{Portinari1998} have the advantage
that they were computed consistently with the classical Padova tracks.
Another merit of the latter is that they distinguish the wind and
core-collapse phases, which is needed when dust ejecta by stars are computed
using the sophisticated model described in Sect.~\ref{sec:dust_evol}.

\begin{figure}
  \includegraphics[width=\columnwidth]{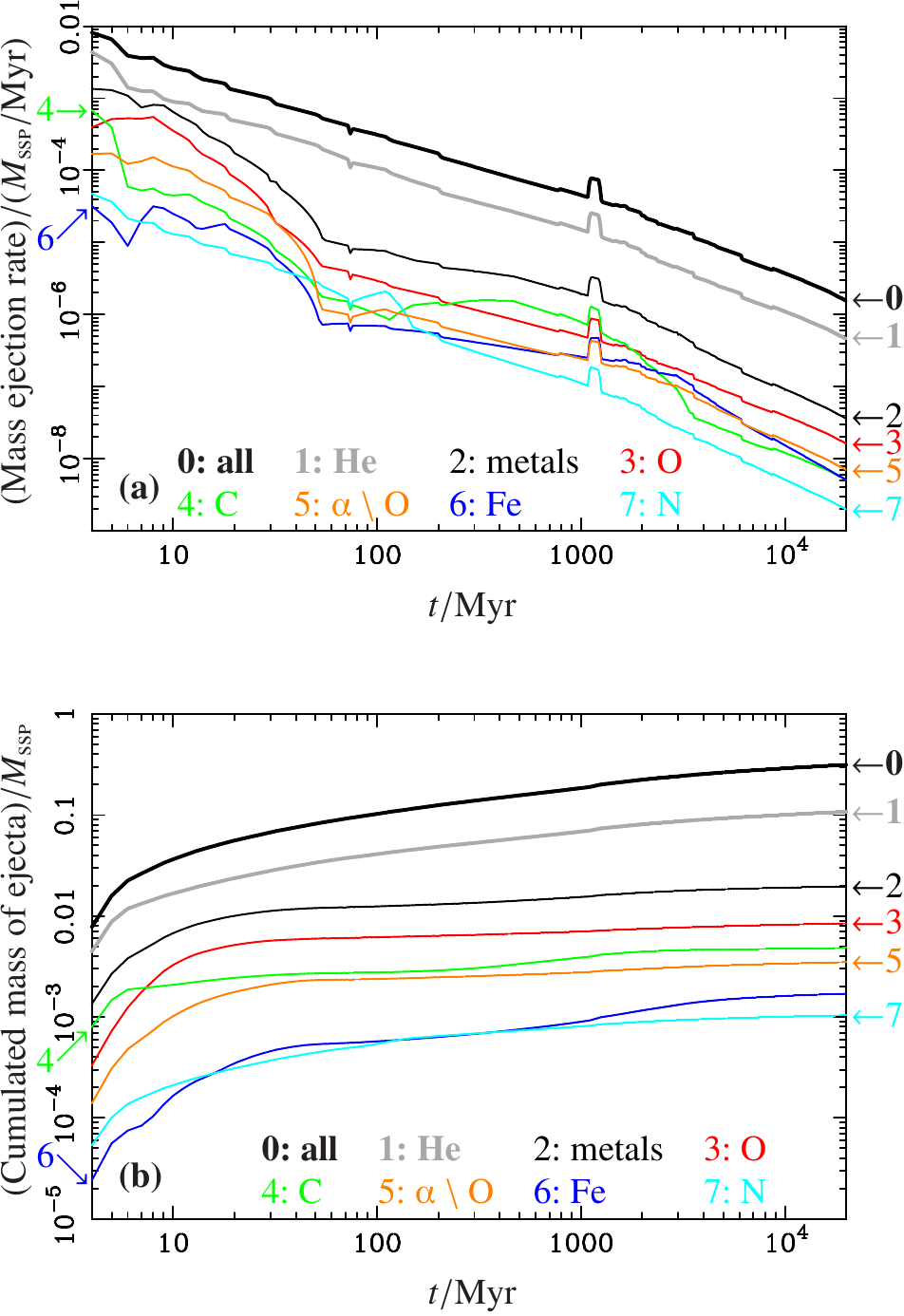}
  \caption{\label{fig:ejecta}%
    Ejecta produced at or until age~$t$ by a single stellar population
    with an initial metallicity $Z = 0.02$ (in mass fraction),
    the \citet{Kroupa1993} IMF and \citet{Portinari1998} yields for massive
    stars; the fraction of close binary stellar systems is $0.05$.\endgraf
    \textbf{(a)}~Ejection rate in mass of ejecta per unit time per unit
    initial mass of the SSP:
    ``all'' denotes the sum on all ejected elements;
    ``metals'', the sum on all metals;
    ``$\alphaup\setminus\mathrm{O}$'', the sum on the main $\alphaup$-elements
    except~oxygen (i.e., Ne, Mg, Si, S and Ca).
    The bump after $1\mult\Gyr$ is due to RGB~stars undergoing the helium flash.
    \endgraf
    \textbf{(b)}~Cumulated mass of ejecta per unit initial mass of the SSP.}
\end{figure}

According to the favorite model, type~Ia supernovae occur in close binaries
where the primary star becomes a CO white dwarf ending in a thermonuclear
explosion.
We use the prescriptions of \citet{Greggio+Renzini} and \citet{Matteucci1986}
to model the number of close binaries and the rate of type~Ia supernovae.
The ejecta produced by the exploding CO~white dwarf are those of model~W7 of
\citet{Thielemann1986}.
As an example, we show in Fig.~\ref{fig:ejecta}
the ejection rate of various
elements by a single stellar population with a metallicity $Z = 0.02$,
and the ejecta cumulated since the birth of the SSP.
\subsubsection{Libraries of stellar spectra}
The main aim of \PEGASE.3 is to model the effects of dust on the spectral
energy distribution of a galaxy.
To do this, the library of stellar spectra must have a large and continuous
wavelength coverage, from the \mbox{far-UV} to the \mbox{near-IR} (at least),
but a high spectral resolution is not required.
The library of stellar spectra used in \PEGASE.3 is made of two blocks:
BaSeL's spectra for stars with an effective temperature
$\Teff < 50\mkern2mu000\mult \mathrm{K}$;
the spectra from \citet{Rauch2003}, rebinned to the wavelengths of BaSeL,
for hotter stars (available only at $[\mathrm{Fe}/\mathrm{H}] \in \{-1, 0\}$).
The emission rate of Lyman continuum photons by stars is computed from these
spectra.

The BaSeL library is based on theoretical spectra, from \citet{Kurucz1979}
mostly, corrected to fit in the \mbox{near-UV}--\mbox{near-IR} domain the
observed colors of stars and star clusters with various metallicities.
Two versions are implemented in the code:
the default one, v2.2 \citep{Lejeune1998}, and v3.1 \citep{Westera2002}%
\footnote{%
  Other libraries of stellar spectra, at higher spectral resolution but
  restricted to the visible, are used in the codes
  \PEGASE-HR (\citealp{LeBorgne2004};
  available at
  \ifdefined\href
  \href{http://www.iap.fr/pegase/}{\texttt{www.iap.fr\slash pegase/}}%
  \else
  \texttt{www.iap.fr\slash pegase/}%
  \fi)
  and \PEGASE-HR2 (near submission).
  The latter covers in particular the wavelength domain observed by \emph{Gaia};
  it is available on request to the \PEGASE\ team members.
}.
\subsection{Star formation history, mass assembly and chemical evolution}
The stellar content of a galaxy is composed of SSPs with various ages and
metallicities.
The unattenuated stellar monochromatic luminosity of a galaxy, at age~$t$
and wavelength~$\lambda$, is
\begin{equation}
  \label{eq:CSP}
  L_\lambda^{\stel,\,\unatt}(t) =
  \int_{t'=0}^t \psi(t-t')\mult L_\lambda^\SSP(t', Z[t-t'])\mult \df t'\,\!,
\end{equation}
where $\psi(t-t')$ is the star formation rate (SFR) at time $t-t'$,
and $Z(t-t')$ is the metallicity of the interstellar medium at that time.

In the code, the star formation rate is computed from the star formation law
the user chose for the scenario.
The parameters determining the infall rate of gas from the reservoirs
onto the galaxy and the outflow rate of matter from the galaxy into the
intergalactic medium \mbox{--~}that is, the mass assembly history of the galaxy~--
are also set by the scenario.
A large variety of scenarios may be built from the trees of parameters
described in the code's documentation.
In particular, one or more instantaneous or extended, overlapping or consecutive
episodes of star formation may occur, and the same holds for infall and outflow.

The first reason why, since \citet{these_Fioc} and \citet{RVF99},
open-box models are preferred in \PEGASE\ is that simpler, closed-box models
predict the presence of a large proportion of low-metallicity stars, which is
detected neither in spiral galaxies \citep{Prantzos+Silk}
\mbox{--~}the ``\mbox{G-dwarf} problem''~-- nor in ellipticals \citep{HW}
and would lead to bluer UV-to-\mbox{near-IR} galaxy SEDs and colors
than what is observed.
Closed-box models also fail to account for the age-metallicity relation
in the Milky Way \citep[see, e.g.,][]{Twarog,Tosi}.
Conversely, \citet{Boissier+Prantzos} and \citet{TCBF} respectively
modeled late-type and early-type galaxies satisfactorily with infall
\citep[see also][]{Sommer-Larsen+2003}.
As emphasized by \citet{Larson72} and \citet{Lynden-Bell},
it is unrealistic to postpone the beginning of star formation
in a model until the complete assembly in the galaxy proper of all its gas.

As regards outflows of matter, whatever their cause (e.g.,
large-scale galactic winds driven by supernovae, as in
\citealp{Mathews+Baker}), they are required to explain the enrichment
of the intergalactic medium and are a possible explanation for the
star formation quenching observed in early-type galaxies
\citep{Pozzetti+2010,Ciesla+2016}.
For a recent review of the many theoretical and observational motivations
to consider exchanges of matter between galaxies and their circumgalactic
environment, see \citet{Tumlinson+2017}.

The chemical evolution of a galaxy is likewise determined by the scenario.
For the evolution of the mass of metals, for instance,
\begin{equation}
  \begin{split}
    \!
    \frac{\df(M_\ISM\mult Z_\ISM)}{\df t}
    &= -(\psi\mult Z_\ISM)(t)\\
    &\mathrel{\phantom{=}}\mathord{}\mskip-\medmuskip
    + \sum_{j=1}^{n_{\mathrm{res}}} \dot M_{\infall,\,j}(t)\mult Z_{\infall,\,j}
    -(\dot M_\outflow\mult Z_\ISM)(t)\\
    &\mathrel{\phantom{=}}\mathord{}\mskip-\medmuskip
    + \int_{t'=0}^t \psi(t-t')\mult \dot M_{\mathrm{ej},\,Z}^\SSP(t', Z[t-t'])\mult \df t'\,\!,
  \end{split}
  \label{eq:metalevolution}
\end{equation}
where $M_\ISM$ and $Z_\ISM$ are the mass of matter and the metallicity in
the ISM, $n_{\mathrm{res}}$ is the number of reservoirs,
$\dot M_{\infall,\,j}$ and $Z_{\infall,\,j}$ are the infall rate from reservoir~$j$
and its metallicity,
$\dot M_\outflow$ is the outflow rate into the intergalactic medium,
and $\dot M_{\mathrm{ej},\,Z}^\SSP$ is the mass ejection rate of metals
by an SSP into the ISM.
The code computes in the same way the evolution of the ISM~abundances of
He, C, N, O, Ne, Mg, Si, S, Ca and Fe.
These equations assume that stellar ejecta are instantaneously and
homogeneously mixed with the ISM once in it, and that the composition
of galactic outflows is the same as that of the ISM.
\section{Properties of grains and dust evolution%
  \label{sec:grains}}
\subsection{Grain sizes and optical properties}
Two families of dust grains are considered.
The first family contains only one species, silicate grains.
The other family, the carbonaceous grains, is subdivided in three species:
graphites, neutral polycyclic aromatic hydrocarbons (PAHs) and ionized PAHs.

In the code, the weights of grain species within their family
and the size distribution of grains in a given species are assumed to be
constant.
The following models of these properties are implemented:
(1)~the ``\textsc{bare\_gr\_s}'' model of \citet{Zubko2004} (by default);
(2)~model number~$7$ in table~1 of \citet{Weingartner2001},
used in \citet{Li2001};
(3)~the outdated model, without PAHs, of \citet{Mathis1977}.

The global optical properties of dust species are computed as a function
of wavelength from the size distributions and from the optical properties
of individual grains given by \citet{Laor1993}, \citet{Draine1984} and
\citet{Li2001}.
The extinction opacity in surface per unit mass (also called the
``mass extinction coefficient'') of dust species~$i$ at wavelength~$\lambda$ is
\begin{equation}
  \label{eq:kappa_ext,i}
  \kappa_{i,\,\lambda}^\extin =
  \int_a \pi\mult  a^2\mult \Bigl(Q_{i,\,\lambda}^\absor[a]
  + Q_{i,\,\lambda}^\scat[a]\Bigr) \mult \frac{\df n_i}{\df a}\mult \df a\,,
\end{equation}
where $\pi\mult  a^2\mult  Q_{i,\,\lambda}^\absor(a)$
(resp. $\pi\mult  a^2\mult  Q_{i,\,\lambda}^\scat(a)$)
is the absorption (resp. scattering) cross-section of a grain with radius~$a$,
and $\df n_i/\df a$ is the number of grains per unit radius per unit mass
of dust.

The overall optical properties of dust at galactic age~$t$ are computed
as a linear combination of the properties of grain species,
weighted by the mass of each species at~$t$.
For instance, the overall extinction opacity of dust is given by
\begin{equation}
  \label{eq:kappa_ext,tot}
  \kappa_\lambda^\extin(t) = \sum_i w_i(t)\mult \kappa_{i,\,\lambda}^\extin\,,
\end{equation}
where $w_i(t)$ is the mass fraction of dust species~$i$ relative to the overall
mass of dust.
Similarly, the overall albedo is given by
\begin{equation}
  \omega_\lambda(t) = \frac{\sum_i w_i(t)\mult \kappa_{i,\,\lambda}^\scat}{
    \kappa_\lambda^\extin(t)}\,,
\end{equation}
and the overall asymmetry parameter by
\begin{equation}
  g_\lambda(t) = \frac{\sum_i w_i(t)\mult \kappa_{i,\,\lambda}^\scat
    \mult g_{i,\,\lambda}}{
    \kappa_\lambda^\extin(t)\mult \omega_\lambda(t)}\,,
\end{equation}
where $g_{i,\,\lambda}$ is the asymmetry parameter of species~$i$.
\subsection{Dust evolution%
  \label{sec:dust_evol}}
While grain size distributions and the relative masses of the three species
of carbonaceous grains are assumed to be constant, the overall masses of
silicate and carbonaceous dust in the ISM evolve with time.
To compute these, we propose two models: a basic one and a more sophisticated
one.
(The default values of the parameters of these models are given in the code's
documentation.)

In the basic model, the mass of dust is simply proportional to the mass of its
constituents in the ISM.
The mass of carbonaceous grains in the ISM at galactic age~$t$ is thus given by
\begin{equation}
  \label{eq:M_carb_formed_anew}
  M_\carbon^\ISM(t) = \delta_\carbon^\ISM \mult  M_{\mathrm{C}}^\ISM(t)\,,
\end{equation}
where $M_{\mathrm{C}}^\ISM$ is the mass of carbon in the ISM (including dust
grains) and $\delta_\carbon^\ISM$ is a constant depletion factor.
(The mass of hydrogen atoms in carbonaceous grains is neglected.)
The mass of silicate grains in the ISM is derived in a similar way:
\begin{equation}
  \label{eq:M_sil_formed_anew}
  M_\silic^\ISM(t) = \delta_\silic^\ISM \mult
  \sum_{i=1}^{n_\silic} {M_i^\ISM(t) \mult  (1 + \OtoSil\mult A_{\mathrm{O}}/A_i)}\,,
\end{equation}
where
$\delta_\silic^\ISM$  is a constant;
the $n_\silic$ elements referred to by the index~$i$ are Mg, Si, S, Ca and Fe;
$M_i^\ISM$ is the mass of element~$i$ in the ISM (including dust grains),
$A_i$ its atomic mass and $A_{\mathrm{O}}$ that of oxygen;
$\OtoSil$ is the number of oxygen atoms in silicate dust per atom of any of
the $n_\silic$ elements.

The sophisticated model was developed by \citeauthor{Dwek1998}
(\citeyear{Dwek1998}; later improved by \citealp{Galliano2008})
and is extensively described in the code's documentation.
It is more physical as it attempts to follow the formation of dust
in the late phases of stellar evolution,
whether in the winds of mass-losing stars or in the ejecta of supernovae,
its destruction in the ISM by the blast waves generated by supernovae,
and the accretion of dust constituents on grains already present in the ISM.
Although it agrees reasonably well with observational data on the dust content
of the Milky Way (MW), it depends on several poorly constrained parameters:
depletion factors of carbonaceous and silicate grains
in either stellar winds or supernova ejecta;
formation efficiency of CO molecules in these environments;
mass of the ISM swept by a single supernova explosion; accretion timescale on grains in the ISM.
For this reason, the results presented in the following were obtained
with the basic model.
\section{Attenuation by dust grains in the diffuse ISM%
  \label{sec:diffuse_attenuation}}
At any wavelength, the attenuation of the stellar emission by dust in the
diffuse interstellar medium (DISM) depends not only on the masses of dust
species and their overall optical properties, but also on the relative
spatial distributions \mbox{--~}the geometry~-- of stars and dust,
and on the viewing angle~$\iota$ toward the galaxy.
We precomputed grids
of the transmittance $\Theta_\lambda = \Latt_\lambda/\Lunatt_\lambda$,
where $\Lunatt_\lambda$ is the unattenuated monochromatic luminosity and
$\Latt_\lambda$ the attenuated one,
for a wide range
of extinction optical depths~$\tau^\extin_\lambda$, albedos $\omega_\lambda$,
asymmetry parameters $g_\lambda$ (from which the scattering angle of photons by
grains is drawn using the probability distribution of \citealp{HG})
and, for spirals, viewing angles.
Monte Carlo simulations of radiative transfer taking scattering into account
and based on the method of virtual interactions \citep{Varosi+Dwek} were used
for this.
Besides the simplistic slab model already implemented in \PEGASE.2,
two geometries are available: one for spiral galaxies, the other for
spheroidal galaxies.
All these grids of transmittance are provided with the code's source files.
\subsection{Spiral galaxies%
  \label{sec:spiral}}
In this geometry, stars are distributed in a disk and a bulge.
The mass density of stars in the disk is modeled as
\begin{equation}
  \mu_\stel^\disk(x, y, z, t) \propto \exp\!\left(-\frac{\rho}{R^\disk_\stel} -
    \frac{\lvert z\rvert}{h^\disk_\stel}\right),
\end{equation}
where $x$, $y$, $z$ are Cartesian coordinates and $\rho = \sqrt{x^2 + y^2}$.
For the mass density of stars in the bulge, we take
\begin{equation}
  \mu_\stel^\bulge(x, y, z, t) \propto \left(\frac{r}{b}\right)^{-q}\mult
  \exp\!\left(-\left[\frac{r}{b}\right]^{1/\nSersic}\right),
\end{equation}
where $r = \sqrt{x^2 + y^2 + z^2}$.
As shown by \cite{LimaNeto99}, the projection of $\mu_\stel^\bulge$ on the
$(x, y)$ plane is very close to a S\'ersic profile of parameter~$\nSersic$
and effective radius  $R^\bulge_\stel$ for appropriate values of $q$ and $b$.

Dust is distributed in a disk, with a mass density
\begin{equation}
  \label{eq:mu_spir_dust}
  \mu_\dust^\spiral(x, y, z, t)
  = \mu_{0,\,\dust}^\spiral(t)\mult \exp\!\left(-\frac{\rho}{R_\dust} -
    \frac{\lvert z\rvert}{h_\dust}\right).
\end{equation}
The central density $\mu_{0,\,\dust}^\spiral$ is fixed by the total
mass of dust in the galaxy, computed from the scenario at age~$t$,
\begin{equation}
  M_\dust^\spiral(t)
  = 4\mult \pi\mult \mu_{0,\,\dust}^\spiral(t)\mult R_\dust^2\mult h_\dust.
\end{equation}
The central column density of dust through the whole galaxy,
computed perpendicularly to the plane of the disk, is then given by
\begin{equation}
  \sigma_{0,\,\dust}^\spiral(t)
  = 2\mult \mu_{0,\,\dust}^\spiral(t)\mult h_\dust\,,
\end{equation}
in mass per unit surface.
The corresponding extinction optical depth is thus
\begin{equation}
  \tau^\extin_\lambda(t)
  = \sigma_{0,\,\dust}^\spiral(t) \mult  \kappa_\lambda^\extin(t)
  = \frac{M_\dust^\spiral(t)}{2\mult \pi\mult R_\dust^2}
  \mult \kappa_\lambda^\extin(t).
\end{equation}

Grids of the transmittances $\Theta_\lambda^\disk$ and $\Theta_\lambda^\bulge$,
through the dust disk, of the light emitted by the stellar disk and bulge
were computed, as a function of $\tau^\extin_\lambda$, $\omega_\lambda$,
$g_\lambda$ and $\iota$, for the geometrical model described
in Table~\ref{tab:fixed_spiral_param}.
Grids of the transmittances $\bar\Theta_\lambda^\disk$ and
$\bar\Theta_\lambda^\bulge$ averaged over all viewing angles
were also produced.

\begin{table}
  \caption{\label{tab:fixed_spiral_param}%
    Geometric parameters used for disks and bulges in the radiative transfer
    Monte Carlo simulations run for spirals.
  }
  \begin{tabular}{ll}
    \hline
    \vrule height 2ex depth 1ex width0pt
    $h^\disk_\stel\mkern-3mu/R^\disk_\stel = 0.1$ & From \citet{Xilouris1999}.
    \\
    \vrule height 2ex depth 1ex width0pt
    $R_\dust/R^\disk_\stel = 1.4$ &
    \\
    \vrule height 2ex depth 1ex width0pt
    $h_\dust/h^\disk_\stel = 0.5$ &
    \\
    \hline
    \vrule height 2ex depth 1ex width0pt
    $\nSersic = 2$ & From \citet{Graham2008}.
    \\
    \vrule height 2ex depth 1ex width0pt
    $R^\bulge_\stel\mkern-3mu/R^\disk_\stel = 0.2$ &
    \\
    \hline
  \end{tabular}
\end{table}

Since the bulge and the disk are distinguished in the code only in their
attenuation but not their evolution, the monochromatic luminosity of the
whole galaxy, as seen from a viewing angle~$\iota$, is given after attenuation
through the diffuse medium by
\begin{align}
  \!
  \Latt_\lambda(t, \iota)
  &= \Theta_\lambda^\disk(t, \iota) \mult L_\lambda^{\disk,\,\unatt}(t)
  + \Theta_\lambda^\bulge(t, \iota)\mult L_\lambda^{\bulge,\,\unatt}(t)
  \notag\\
  &= \left(
    [1-\BoverT]\mult \Theta_\lambda^\disk[t, \iota]
    +\BoverT\mult \Theta_\lambda^\bulge[t, \iota]\right)\mult\Lunatt_\lambda(t)\,,
  \label{eq:theta_spir}
\end{align}
where $L_\lambda^{\disk,\,\unatt}$ and $L_\lambda^{\bulge,\,\unatt}$
are the unattenuated luminosities of the disk and the bulge,
and $\BoverT$ is the constant bulge-to-total mass ratio.

A final parameter is required to compute the attenuation by the diffuse medium,
but contrary to the quantities listed in Table~\ref{tab:fixed_spiral_param},
this parameter must involve absolute sizes and masses.
The code uses $M_\system^\spiral/R_\dust^2$, where $M_\system^\spiral$ is the mass
of the system for a spiral geometry.

The default values for this parameter and for the bulge-to-total mass
ratio are $M_\system^\spiral/R_\dust^2 = 2915\mult M_\odot/\parsec^2$
and $\BoverT = 1/7$.
We derived them from the Milky Way model in table~I-2
of \citet{Binney2008}.
Contrary to the values listed in Table~\ref{tab:fixed_spiral_param},
the user may change them.
\subsection{Spheroidal galaxies%
  \label{sec:spher}}
The spatial distribution of stars in spheroidal galaxies
is modeled with a King profile:
the mass density of stars is given by
\begin{equation}
  \label{eq:mu_sph_stell}
  \mu_\stel^\spher(x, y, z, t) =
  \begin{cases}
    \mu_{0,\,\stel}^\spher(t)\mult \Bigl(1 + [r/\Rcore]^2\Bigr)^{-3/2}
    &  \text{for } r \leqslant \Rtrunc\,,
    \\
    0
    & \text{for } r > \Rtrunc\,,
  \end{cases}
\end{equation}
where $\Rcore$ is the core radius and $\Rtrunc$ is the truncation radius
\citep{Tsai1995}.

For dust, we follow \citet{Froehlich} and model
the mass density as
\begin{align}
  \label{eq:mu_sph_dust}
  \!
  \mu_\dust^\spher(x, y, z, t) &\propto \Bigl(\mu_\stel^\spher[x, y, z, t]\Bigr)^{\nFroehlich}
  \notag\\
  &= \begin{cases} \mu_{0,\,\dust}^\spher(t)\mult \Bigl(1 + [r/\Rcore]^2\Bigr)^{-3\mult \nFroehlich/2}
    &\text{for } r \leqslant \Rtrunc\,,
    \\
    0
    &\text{for } r > \Rtrunc.
  \end{cases}
\end{align}
The central density $\mu_{0,\,\dust}^\spher$ is fixed by the total
mass of dust in the galaxy, computed from the scenario at age~$t$,
\begin{equation}
  M_\dust^\spher(t)
  = 4\mult \pi\mult \Rcore^3\mult \mu_{0,\,\dust}^\spher(t)\mult
  \int_{s=0}^{\Rtrunc/\Rcore} {s^2\mult \bigl(1+s^2\bigr)^{-3\mult \nFroehlich/2}\mult \df s}\,,
\end{equation}
where $s=r/\Rcore$.
The central column density of dust through the whole galaxy,
in mass per unit surface, is then given by
\begin{equation}
  \sigma_{0,\,\dust}^\spher(t)
  = 2\mult \Rcore\mult \mu_{0,\,\dust}^\spher(t)\mult
  \int_{s=0}^{\Rtrunc/\Rcore} {\bigl(1+s^2\bigr)^{-3\mult \nFroehlich/2}\mult \df s}.
\end{equation}
The corresponding extinction optical depth is therefore
\begin{equation}
  \label{eq:tau_lambda_sph}
  \tau^\extin_\lambda(t)
  = \frac{\int_{s=0}^{\Rtrunc/\Rcore} {\bigl(1+s^2\bigr)^{-3\mult \nFroehlich/2}\mult \df s}}{2\mult \pi\mult
    \int_{s=0}^{\Rtrunc/\Rcore} {s^2\mult \bigl(1+s^2\bigr)^{-3\mult \nFroehlich/2}\mult \df s}}
  \mult \frac{M_\dust^\spher(t)}{\Rcore^2}\mult \kappa_\lambda^\extin(t).
\end{equation}

A grid of the transmittance $\Theta^\spher_\lambda$ was computed for
spheroidal galaxies as a function of $\tau^\extin_\lambda$, $\omega_\lambda$
and $g_\lambda$, taking $\nFroehlich = 1/2$, as recommended by \citet{Tsai1995},
and using the value $\Rtrunc/\Rcore = 397.92$ of their model~\emph{b}.
One has then
\begin{equation}
  \label{eq:tau_lambda_sph_num}
  \tau^\extin_\lambda(t)
  = 7.587\times10^{-5}\mult \frac{M_\dust^\spher(t)}{\Rcore^2}
  \mult \kappa_\lambda^\extin(t).
\end{equation}

As with spirals, a final parameter, involving absolute sizes and masses,
is required to compute the attenuation by the diffuse medium.
The code uses $M_\system^\spher/\Rcore^2$, where $M_\system^\spher$ is the mass
of the system for a spheroidal geometry.
The default value of this ratio,
$M_\system^\spher/\Rcore^2 = 7.73\times10^6\mult M_\odot\mult \parsec^{-2}$,
was computed from the values in model~\emph{b} of \citet{Tsai1995}, taking the
current mass of stars in this model for $M_\system^\spher$.
The user may change it, contrary to the values of $\nFroehlich$ and
$\Rtrunc/\Rcore$ used in the simulations.
\section{Emission of light by dust grains in the diffuse ISM%
  \label{sec:diffuse_emission}}
\subsection{Mean radiation field}
The emission of a dust grain at a position~$\vec r$ in the galaxy and age~$t$
depends on its optical properties, related to its species~$i$ and its size~$a$,
and on the local interstellar radiation field (ISRF)~$u_\lambda(\vec r, t)$
at~$\vec r$ and~$t$.
Computing the ISRF at each point in the galaxy and the resulting grain
emission would however be excessively time-consuming.
Instead of this, the code uses the mean radiation field in the diffuse
ISM, $\langle u_\lambda\rangle(t)$, which is calculated as follows:
\begin{align}
  \!
  \Lunatt_\lambda(t) - \meanLatt_\lambda(t)
  \hskip-4em\relax&
  \notag\\
  &
  = \sum_i\iiint_{\vec r}\int_a c\mult  u_\lambda(\vec r, t)
  \mult \pi\mult  a^2\mult  Q_{i,\,\lambda}^\absor(a)
  \mult \frac{\df n_i}{\df a}\mult \mu_i(\vec r, t)\mult \df^3\vec r\mult \df a
  \notag\\
  &
  = c\mult \langle u_\lambda\rangle(t)\mult \kappa^\absor_\lambda(t)
  \mult M_\dust(t)\,,
  \label{eq:mean_ISRF}
\end{align}
where
$\Lunatt_\lambda$ is the unattenuated monochromatic luminosity of the galaxy,
and $\meanLatt_\lambda$ the attenuated luminosity averaged over all viewing
angles;
$\sum_i$ is a sum on all dust species,
$\iiint_{\vec r}$ an integral on the whole galaxy
and $\int_a$ an integral on all grain sizes;
$c$ is the speed of light;
$Q_{i,\,\lambda}^\absor(a)$ and $\df n_i/\df a$ have already been defined
after Eq.~\eqref{eq:kappa_ext,i};
$\kappa^\absor_\lambda$ is the overall absorption opacity,
computed in the same way as the extinction opacity $\kappa^\extin_\lambda$
from Eqs.~\eqref{eq:kappa_ext,i} and~\eqref{eq:kappa_ext,tot}, but without
the scattering term;
$\mu_i(\vec r, t)$ is the spatial density of dust species~$i$ in the ISM
(not the inner density of a grain) at $\vec r$ and~$t$ and is computed as
$\mu_i(\vec r, t) = w_i(t)\mult \mu_\dust(\vec r, t)$,
where  $w_i$ has been defined after Eq.~\eqref{eq:kappa_ext,tot}, and
the spatial density of dust $\mu_\dust$ is given by Eq.~\eqref{eq:mu_spir_dust}
or~\eqref{eq:mu_sph_dust}, depending on the geometry of the galaxy;
$M_\dust$ is the mass of dust in the diffuse medium and is computed as
detailed in Sect.~\ref{sec:dust_evol}.

Finally,
\begin{equation}
  \label{eq:mean_ISRF2}
  \langle u_\lambda\rangle(t)
  = \frac{\bigl(1-\bar\Theta_\lambda[t]\bigr)\mult\Lunatt_\lambda(t)}{
    c\mult \kappa^\absor_\lambda(t)\mult M_\dust(t)}\,,
\end{equation}
where $\bar\Theta_\lambda$ is the transmittance through the diffuse ISM of the
light emitted by the galaxy in all directions.
(In the case of spirals,
$\bar\Theta_\lambda = (1-\BoverT)\mult \bar\Theta_\lambda^\disk
+ \BoverT\mult \bar\Theta_\lambda^\bulge$;
see Eq.~\eqref{eq:theta_spir}.
For spheroidals, $\bar\Theta_\lambda = \Theta^\spher_\lambda$.)
Although using the mean interstellar radiation field instead of the local ISRF
at each point in the galaxy automatically respects the conservation of energy,
it is clear that this narrows the probability distribution of
dust grain temperatures and the dust emission spectrum.
\subsection{Stochastic heating%
  \label{sec:stoch_heat}}
The probability distribution $\df P/\df T$ of the temperatures $T$ of dust
grains in the diffuse ISM is then computed from the mean interstellar radiation
field $\langle u_\lambda\rangle$, taking into account stochastic heating.
The procedure described in \citet{Guhathakurta1989}, which assumes
that the cooling of grains is continuous, was adopted:
only the heating due to photons is considered, so collisions and
grain sublimation are neglected.
The internal energies of grains are calculated from the prescriptions in
\citet{Draine2001} and \citet{Li2001}.

\begin{figure}
  \includegraphics[width=\columnwidth]{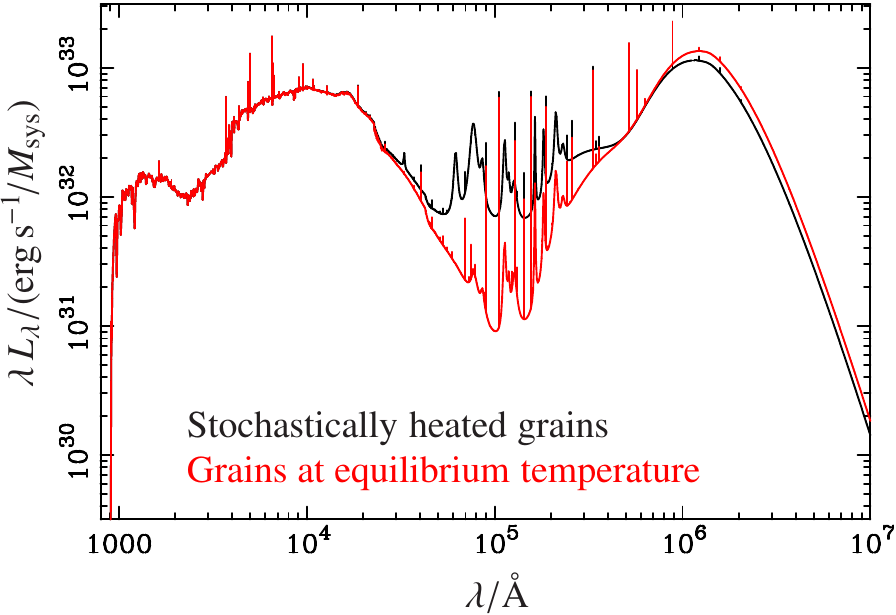}
  \caption{\label{fig:stochastic}%
    Effects of stochastic heating on the SED of the $13\mult\Gyr$-old Milky
    Way model (see Sect.~\ref{sec:MW_model}).
    All the SEDs shown hereafter were averaged over all viewing angles.
    For emission lines, only the peak was plotted;
    its height was computed assuming a Gaussian profile
    with a full width at half maximum of $160\mult\mathrm{km/s}$
    \citep{Mocz+2012}.
    Black line: stochastically heated grains.
    Red line: all grains were assumed to have reached their equilibrium
    temperature.
  }
\end{figure}

The overall monochromatic luminosity of dust is given by
\begin{equation}
  \begin{split}
    \!
    L^\dust_\lambda(t) =
    M_\dust(t) \mult
    \sum_i w_i(t)\mult \int_a
    & 4\mult \pi\mult a^2\mult Q_i^\absor(a)\mult \frac{\df n_i}{\df a}
    \\
    & \times\int_T \frac{\df P_i}{\df T}(a, T, t)
    \mult X_\lambda(T)\mult \df T\mult \df a\,,
  \end{split}
\end{equation}
where
\begin{equation}
  X_\lambda(T) = \pi\mult \frac{2\mult h\mult c^2}{
    \lambda^5\mult \Bigl(\exp\Bigl[\frac{h\mult c}{
      \lambda\mult k_{\mathrm{B}}\mult T}\Bigr]-1\Bigr)}
\end{equation}
is the spectral exitance of a blackbody, and $h$ and $k_{\mathrm{B}}$ are the
Planck and Boltzmann constants.
Figure~\ref{fig:stochastic} illustrates the effects of stochastic heating on
the SED of the $13\mult\Gyr$-old Milky Way model described in
Sect.~\ref{sec:MW_model}.

By default, the self-absorption by dust grains of dust-emitted light
is neglected.
So, if one forgets star-forming clouds for the moment, the luminosity of
a galaxy observed from an angle~$\iota$ would just be
$\Theta_\lambda(t, \iota)\mult\Lunatt_\lambda(t) + L^\dust_\lambda(t)$ at time $t$.
To assess the effects of self-absorption, a crude modeling of this phenomenon
was however implemented in the code (see the documentation).

\section{Star-forming regions and nebular emission%
  \label{sec:cloud}}
The code processes the clouds of interstellar matter surrounding star-forming
regions separately from the diffuse interstellar medium (DISM).
Star-forming clouds are modeled as homogeneous spherical shells of gas and
dust, at the center of each of which resides a point-like cluster of young
stars.
The size of clouds is considered as negligible compared to that of the
diffuse medium in which they are embedded;
the emission emerging from any star-forming cloud is therefore subjected to
the same processing by the DISM as that of the old stars scattered through
the latter.

Because of the impact of massive stars on their surrounding, either through
their intense radiation, stellar winds or supernova explosions, the life
duration of star-forming clouds is a few million years at most.
Young stars may also escape from their birth cloud before it is destroyed.
The fraction $\varphi(t')$ of stars aged~$t'$ still in their parent cloud
is modeled as
\begin{equation}
  \label{eq:cloud}
  \varphi(t') = \varphi_0\mult (1 - t'/\theta)^\beta\,\!,
\end{equation}
where $\varphi_0$, $\theta$ and $\beta$ are constant parameters;
$\varphi$ may also represent the covering factor of the cluster by the cloud,
in which case a fraction $1-\varphi$ of the photons emitted by the cluster
leak directly into the diffuse medium.

At any galactic age~$t$, the code computes from the star formation rate
$\psi(t-t')$ the number of clusters with an age $t' \leqslant \theta$
and the emission rate of Lyman continuum photons produced by each cluster,
taking a typical initial stellar mass $M_\cluster$ for the clusters;
this mass and the parameters $\varphi_0$, $\theta$ and $\beta$ mentioned above
may be changed by the user.

\subsection{Nebular emission from star-forming clouds.
  Dust effects on their SED}

\subsubsection{Nebular emission in the dust-free case%
  \label{sec:neb_em_clouds}}
The modeling of nebular emission implemented in previous versions of
\PEGASE\ dated back, for most of the emission lines, to \citet{GRV1987}.
The intensities of these lines were taken from \citet{Stasinska1984}, and
only UV-to-\mbox{near-IR} lines at solar metallicity were considered.
An upgrade was clearly needed.

We therefore computed a grid of models of dust-free \HII~regions with
version c17.01 of the code \textsc{Cloudy} \citep{Cloudy} as a function
of the metallicity of the ISM, $Z_\model$, and of the number rate of
Lyman continuum (\LymCont) photons emitted by the central ionizing source,
$\NLC_\model$.
The values considered ranged from $\mathord{\approx}\,0$ to $0.1$ for $Z_\model$,
and from $10^{46}$ to $10^{53}\mult\mathrm{s}^{-1}$ for $\NLC_\model$.

The geometry adopted in these computations was radiation-bounded, spherical
and with an inner cavity of radius $\rCav = 1\mult\parsec$ centered on
the ionizing source.
The filling and covering factors were set to~$1$ (their default value
in \textsc{Cloudy}).
The number density $\nH$ of hydrogen atoms, be they neutral, ionized,
isolated or in molecules, was assumed to be constant throughout the cloud.
We took $\nH = 10^2\mult\mathrm{cm}^{-3}$.

In addition to the amplitude of the ionizing radiation, characterized
here by $\NLC_\model$, \textsc{Cloudy} needs its spectral shape.
Because the exact shape only has a secondary impact on nebular emission,
we did not use that of the star cluster at age~$t'$:
this would have been very inconvenient as it would have required
to precompute a huge grid
of models for all ages and possible stellar initial mass functions.
Instead, we took
the unattenuated stellar emission produced by a constant
star formation rate at fixed metallicity~$Z_\model$ and at an age
$\mathord{\ga}\,10\mult\Myr$.
The reason for this is that the number rate of \LymCont~photons emitted
by a single stellar population, $\NLC_\SSP(t')$, drops very rapidly
at $t' > 4\mult\Myr$;
so, the SED produced by a constant star formation rate does almost not evolve
in the Lyman continuum at any age larger than $\mathord{\approx}\,10\mult\Myr$.
The resulting shape is a time-average of the ionizing SEDs, weighted by
$\NLC_\SSP(t')$;
it should therefore be representative of the typical ionizing SED
at the metallicity considered.

By default, \textsc{Cloudy} stops its calculations at the distance from the
central source where the electronic temperature falls below
$4000\mult\mathrm{K}$.
This criterion is not appropriate for low values of $\NLC_\model$ or high
values of $Z_\model$, as already mentioned in \citet[sect.~3.4.4]{thesis_Moy}%
\footnote{This Ph.D.\ thesis is written in French;
  significant parts of it are available in English in \citet{article_Moy}.}
and \citet{Gutkin+2016}, because this event occurs then well inside the
\HII~region:
a large number of \LymCont~photons are not absorbed there yet,
and the nebular emission predicted by the code with this criterion
would therefore be unreliable.
We instead required that \textsc{Cloudy} computes the state of the region
up to the distance where the number density of free protons drops below
$10^{-2}\mult \nH$.

For each model in the grid, we extracted from \textsc{Cloudy}'s outputs the
integrated luminosity of emergent emission lines, out of which we selected
all those brighter than $10^{-4}~\mathord{\times}$ the luminosity
of Lyman~$\alphaup$ in at least one of the models of the grid;
this resulted in a final set of $\mathord{\approx}\,400$~lines, including many
IR~lines which were absent from \PEGASE.2.
We also extracted the SED of the pure nebular continuum, uncontaminated by
emission lines, and the value of the Str\"omgren radius of the model,
$\rSz^\model$, defined as the distance to the cluster where the fraction
of neutral hydrogen to all forms of hydrogen reaches $1/2$.

For each star cluster, \PEGASE.3 computes, 
in the dust-free case, the number rate
of \LymCont~photons it emits at age~$t'$ which are absorbed by gas
in the birth cloud as
\begin{equation}
  \label{eq:NLC_cloud}
  \NLC_\birthCloud(t')
  = \varphi(t')\mult M_\cluster\mult \left(\frac{\NLC_\SSP[t']}{M_\SSP}\right)\!\,,
\end{equation}
where $(\NLC_\SSP[t']/M_\SSP)$ is the number rate of \LymCont~photons emitted
at age~$t'$ by an SSP, per unit mass of the SSP.
It then interpolates in the $\{\NLC_\model\} \times \{Z_\model\}$ space
of cloud models at the point $(\NLC_\birthCloud[t'], Z[t])$ to compute,
in the dust-free case, the Str\"omgren radius $\rSz$ of the \HII~region
created in the cloud by the cluster, as well as the luminosities of the
emission lines and of the nebular continuum produced there.
More precisely, the code interpolates the nebular luminosities normalized to
the \LymCont~luminosity (i.e., integrated over the Lyman continuum)
of the model ionizing source, and scales these normalized luminosities
by the fraction of the \LymCont~luminosity radiated by the cluster at
age~$t'$ which is absorbed by gas in the birth cloud (this fraction is
computed in the same way as the quantity $\NLC_\birthCloud(t')$ in
Eq.~\eqref{eq:NLC_cloud}).
\subsubsection{Dust effects}
The \HII~region surrounding a star cluster is also filled with dust which,
because of the competition between gas and grains to absorb Lyman continuum
photons, reduces the radius of the region from $\rSz$ to $\rSd$.
We assume that all the \LymCont~photons emitted by stars still in their
parent cloud (or the photons not leaking directly in the diffuse medium,
if $\varphi$ is interpreted as the covering factor of the cluster)
are absorbed, either by gas or by dust, in the  \HII~region of the cloud.
The volume reduction factor $\Upsilon$ is computed in a way similar
to that in sect.~5.1.c of \citet{Spitzer}.
Because the calculations in this book do no include any central cavity
whereas \textsc{Cloudy} requires one, we had to adapt them;
detailed explanations are provided in App.~\ref{sec:cavity}.
The intensities, when dust is present, of the nebular continuum and emission
lines emerging from the cloud's \HII~region are then computed as
$\Upsilon~\mathord{\times}$ their values in the dust-free case,
while a fraction $1-\Upsilon$ of \LymCont~photons heat dust grains in this
region.

However, because the Lyman~$\alphaup$ line is resonant and has thus a large
scattering optical depth in hydrogen, we assume that, as soon as
any dust is present, all the photons emitted
in this emission line are absorbed by grains, either in the cloud's
\HII~region or in the neutral, \HI~region just surrounding it.
The absorption of other emission lines and of the nebular continuum
by dust in the \HII\ and \HI~regions of clouds is neglected.

Grains in the cloud's \HII~region also absorb a fraction of the non-ionizing
photons emitted by the cluster.
Because scattering by grains is mainly forward at the ultraviolet
wavelengths where the cluster emits most of its light,
we may estimate the fraction of stellar photons emerging from the \HII~region,
at a wavelength~$\lambda$ longwards of the Lyman limit, as
\begin{equation}
  \Theta_\lambda^{\birthCloud}
  \approx \exp\Bigl(-\kappa_\lambda^\absor[t]\mult[\rSd -\rCav]\Bigr).
\end{equation}

\begin{figure}
  \includegraphics[width=\columnwidth]{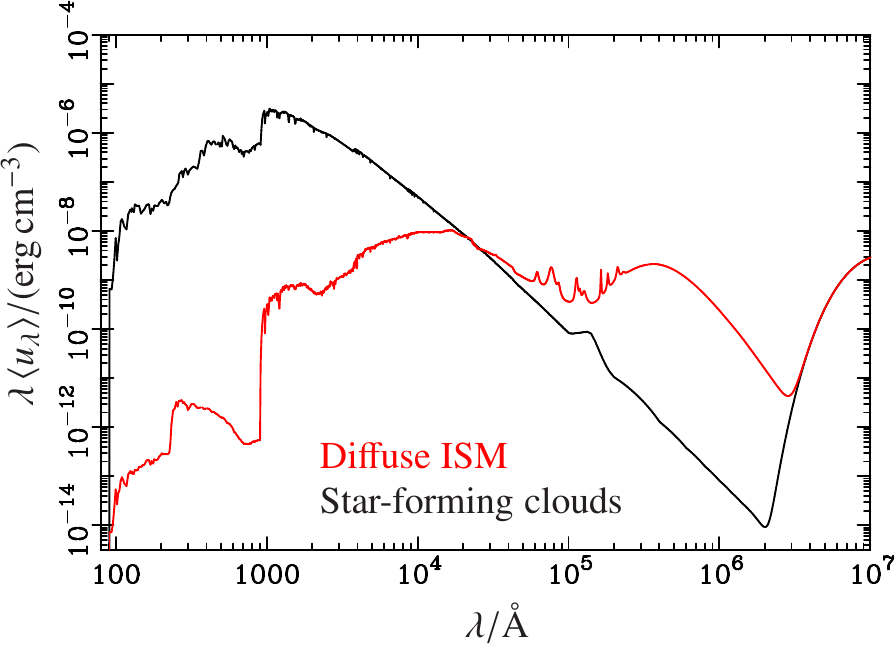}
  \caption{\label{fig:radiation_field}%
    Mean radiation field of the $13\mult\Gyr$-old Milky Way model
    in star-forming clouds (black line) and in the diffuse interstellar medium
    (red line).
    The quantity $\langle u_\lambda\rangle$ is the energy density.
    The bump at long wavelengths is the cosmic microwave background.}
\end{figure}

The code finally determines the interstellar radiation field averaged over all
star-forming clouds, using Eq.~\eqref{eq:mean_ISRF2} with appropriate
substitutions, and computes the emission of dust grains in clouds in the
same way as explained in Sect.~\ref{sec:stoch_heat}.
Except for Lyman~$\alphaup$,
the very unconstrained absorption of the stellar and nebular emission
by dust in the \HI~regions of clouds is not taken into account.
As an example, we compare in Fig.~\ref{fig:radiation_field} the mean radiation
field in the diffuse ISM and in star-forming clouds for the Milky Way model
described in Sect.~\ref{sec:MW_model}.
\subsection{Nebular emission by the diffuse ISM}
A fraction $1-\varphi(t')$ of the Lyman continuum photons emitted
by a single stellar population with an age~$t'$ are produced in the diffuse
interstellar medium (or have leaked into it, if $\varphi$ is interpreted
as the covering factor of the cluster by the birth cloud).
We assume that all these photons are absorbed in the DISM.
The spherical geometry adopted to model star-forming clouds is obviously
not suitable to compute the nebular emission of the diffuse medium.
However, because the ionization parameter~$U$ is the main determinant
(besides metallicity) of the intensity of emission lines and of the nebular
continuum \citep{Davidson77}, we may reuse the results obtained for dust-free
clouds and take, as a proxy and with some appropriate scaling, the nebular
emission of a cloud with a mean value of $U$ equal to that of the DISM.
A typical value of the latter is $U_\DISM \approx 10^{-3.5}$, according to
\citet{U_DISM} \citep[see also][]{Dopita+2006}.

So, for each model of cloud in the grid described in
Sect.~\ref{sec:neb_em_clouds}, we computed the average
$\langle U_\model\rangle$ of the ionization parameter over the volume of the
\HII~spherical shell surrounding the central cluster
\citep{Charlot+Longhetti2001} as follows:
at a distance $r$ from the central source, the ionization parameter is
\begin{equation}
  U_\model(r) = \frac{\NLC_\model}{4\mult \pi\mult r^2\mult \nH\mult c}\,,
\end{equation}
so
\begin{equation}
  \langle U_\model\rangle = \frac{3\mult \NLC_\model}{4\mult \pi\mult \nH\mult c}
  \mult \frac{\rSz^\model-\rCav}{\Bigl(\rSz^\model\Bigr)^3-\rCav^3}.
\end{equation}

\PEGASE.3 interpolates then in the $\{\langle U_\model\rangle\}\times\{Z_\model\}$
space of cloud models at the point $(U_\DISM, Z[t])$ to compute the nebular
emission produced by a cloud in nearly the same state as the diffuse medium
at age~$t$;
this typically corresponds to a cloud model with
$N_\model \approx 10^{47}\mult\mathrm{s}^{-1}$.
Finally, to obtain the overall nebular emission of the DISM, the code scales
the luminosities of the emission lines and of the nebular continuum
emerging from the ``interpolated'' cloud by the ratio of the \LymCont~luminosity
radiated at time~$t$ by stars in the diffuse medium to that of the ionizing
source of the cloud.

As with star-forming clouds, we assume that all the photons emitted
in the Lyman~$\alphaup$ line by the DISM are absorbed by grains as soon as
any dust is present.
The other emission lines and the nebular continuum produced in the diffuse
medium are attenuated in the same way as stars in the latter.

\section{Main outputs%
  \label{sec:outputs}}
For each scenario, the code provides a large number of outputs
as a function of galactic age:
masses and SEDs of various components, chemical abundances,
star formation and supernova rates.
We refer to the code's documentation for a detailed list of all the quantities.
In the following, we present some outputs obtained for a scenario
fitted to observed colors of nearby spiral galaxies of Sbc Hubble type.
As this is the most likely type for the Milky Way
\citep{deVaucouleurs+Pence,Hodge83}, we also constrained this scenario using
observed properties of the MW's local ISM and call it the ``MW model''.
\subsection {Abundances of metallic elements%
  \label{sec:MW_model}}
\begin{figure}
  \includegraphics[width=\columnwidth]{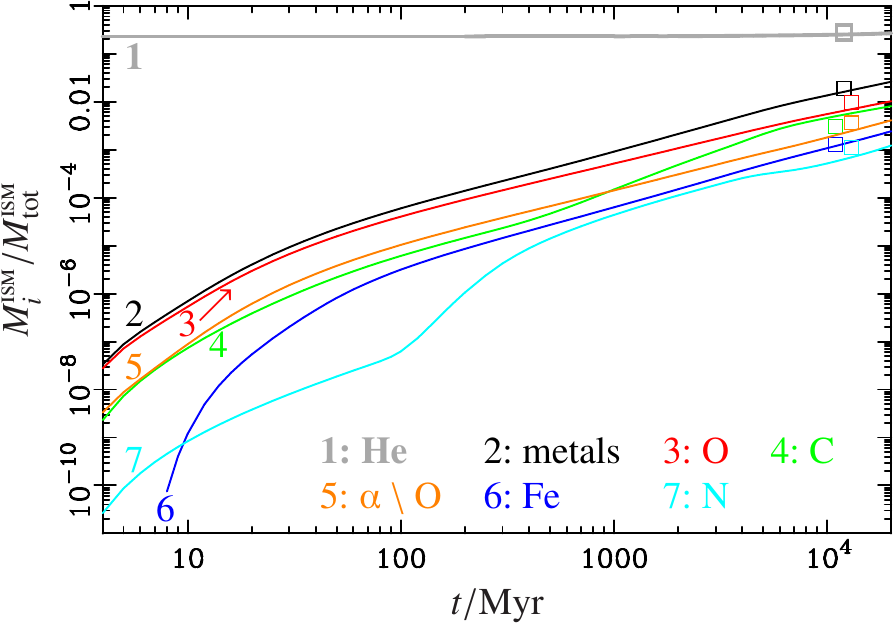}
  \caption{\label{fig:MetalAbondances}%
    Evolution of ISM abundances as a function of galactic age~$t$
    for the Milky Way model (lines):
    $M_i^\ISM$ is the mass in the ISM of the element (or elements) referred
    to by the index~$i$, and $M_\total^\ISM$ is the total mass of the ISM;
    ``metals'' denotes the sum on all metals, and
    ``$\alphaup\setminus\mathrm{O}$'', the sum on Ne, Mg, Si, S and Ca.
    Squares are the corresponding observed abundances of \citet{Anders1989}
    (shuffled around $13\mult\Gyr$ for the sake of clarity).}
\end{figure}

Figure~\ref{fig:MetalAbondances} shows the evolution of the ISM abundances of
several elements for the MW model.
These abundances are compared at $\mathord{\approx}\,13\mult\Gyr$ with
observed abundances in the solar neighborhood from \citet{Anders1989}.
The scenario used for the MW model was the following:
the galaxy assembles by infall at an exponentially decreasing rate,
with a timescale of $6\mult\Gyr$;
there are no galactic outflows;
the star formation rate is proportional to the current mass of the ISM,
with an efficiency of $(6\mult \Gyr)^{-1}$;
the IMF is from \citet{Kroupa1993};
the chemical yields by massive stars are from \citet{Portinari1998};
the values adopted to model star-forming clouds are
$M_\cluster = 10^4\mult M_\odot$, $\varphi_0 = 1$, $\theta = 10\mult\Myr$
and $\beta = 1$.
\subsection {SEDs and colors%
  \label{sec:colors_MW}}
Galaxy SEDs are computed on the wavelength domain defined jointly
by the library of stellar spectra and the model of grains,
continuously from the \mbox{far-UV} to the \mbox{far-IR} and submm.
The spectral resolution decreases from $\lambda/\Delta\lambda \approx 100$
in the \mbox{far-UV} to $\mathord{\approx}\,20$ in the submm.

More than $60$~calibrated filter passbands are provided with the code.
Other filters may easily be added to this list, as long as the reference
spectrum used for the calibration (e.g., Vega, AB) and the
passband transmission (whether in energy or in number of photons) are known.
Magnitudes are calculated from the galaxy SED through all these
passbands.
The user may freely select the magnitudes, luminosities, colors
and line equivalent widths printed by the code.

\begin{figure}
  \includegraphics[width=\columnwidth]{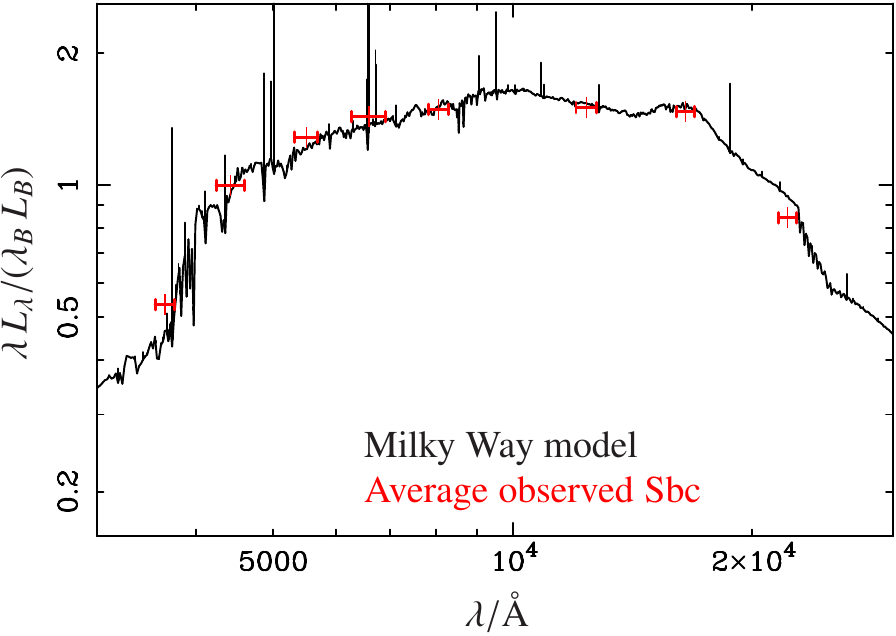}
  \caption{\label{fig:SbcSED}%
    \mbox{Near-UV}-to-\mbox{near-IR} spectral energy distribution
    of the $13\mult\Gyr$-old Milky Way model (black line), compared with the
    mean observed fluxes of local Sbc galaxies (red crosses) in the
    $U$, $B$, $V$, $R_{\mathrm{c}}$, $I_{\mathrm{c}}$, $J$, $H$ and $K$ passbands
    (see Sects.~\ref{sec:MW_model} and~\ref{sec:colors_MW} for details).
    All fluxes were normalized to the $B$ band.
    The red horizontal segments correspond to the width of the passbands.
    See also Table~\ref{tab:Sbc}.}
\end{figure}

As an illustration, we compare, in Table~\ref{tab:Sbc} and
Fig.~\ref{fig:SbcSED}, the \mbox{near-UV}-to-\mbox{near-IR} colors and
spectrum of the Milky Way model at $t = 13\mult\Gyr$ with typical total
observed colors of nearby Sbc~galaxies.
The model spectrum was computed with the BaSeL-2.2 library of stellar
spectra and averaged over all viewing angles.
The observed $B-H$, $J-H$ and $H-K$ colors were compiled by \citet{Fioc1999b},
corrected for aperture and redshift, and gathered in eight types covering
the whole Hubble sequence, one of which being the Sbc type discussed here;
the authors also averaged these colors within each of these types according
to a procedure (described in their paper) taking into account uncertainties
and the intrinsic scatter.
The $U-B$ and $B-V$ (resp.\ $V-R_{\mathrm{c}}$ and $V-I_{\mathrm{c}}$) mean
observed colors of the same morphological types were computed from the data
in \citet{Vaucouleurs1991} (resp.\ \citealp{Buta1995});
we submitted them to the same averaging procedure than $B-H$ and
\mbox{near-IR} colors.
\begin{table}
  \caption{\label{tab:Sbc}%
    Comparison of \mbox{near-UV}-to-\mbox{near-IR} colors of the MW model
    at $t = 13\mult\Gyr$ with typical total observed colors of nearby
    Sbc~galaxies computed by \citet{Fioc1999b}.
    (See Sects.~\ref{sec:MW_model} and~\ref{sec:colors_MW} for details.)
  }
  \begin{tabular}{ccc}
    \hline
    Color & Average observed Sbc & MW model \\
    \hline
    $B-V$  & $0.63
    $ & $0.634$ \\
    $U-B$  & $0.04
    $ & $0.056$ \\
    $V-R_{\mathrm{c}}$ & $0.48
    $ & $0.509$  \\
    $V-I_{\mathrm{c}}$ & $1.01
    $ & $1.088$  \\
    $J-H$  & $0.77
    $ & $0.750$ \\
    $H-K$  & $0.22
    $ & $0.307$ \\
    $B-H$  & $3.33
    $ & $3.408$  \\
    \hline
  \end{tabular}
\end{table}
\subsection{Evolution of the SED and of galactic components}
\begin{figure}
  \includegraphics[width=\columnwidth]{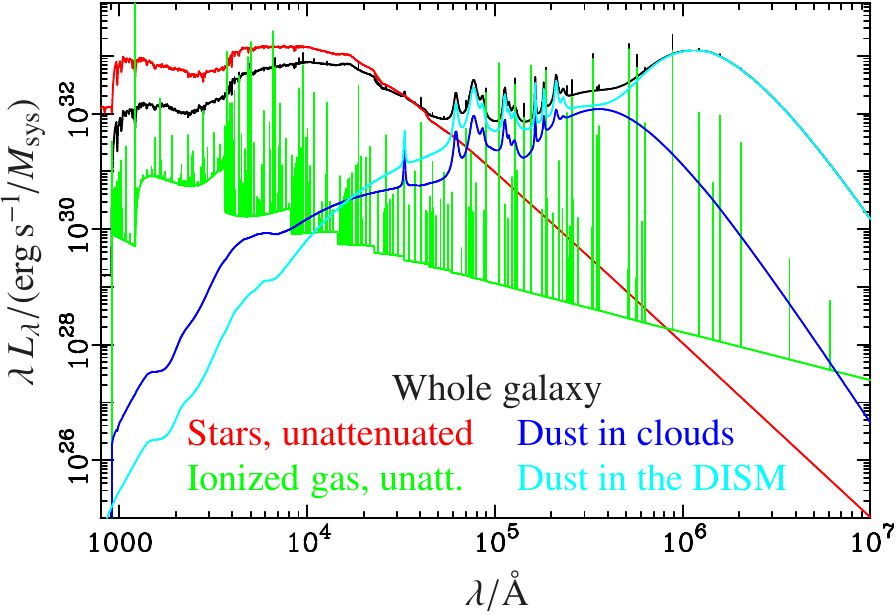}
  \caption{\label{fig:Composantes_13Gyr}%
    Spectral energy distribution of the $13\mult\Gyr$-old Milky Way model
    (black line) and of its components:
    unattenuated nebular continuum and lines produced by the ionized gas
    (green line);
    dust emission from star-forming clouds (dark blue line);
    dust emission from the diffuse ISM (light blue line).
    The unattenuated stellar continuum is plotted in red.}
\end{figure}

\begin{figure}
  \includegraphics[width=\columnwidth]{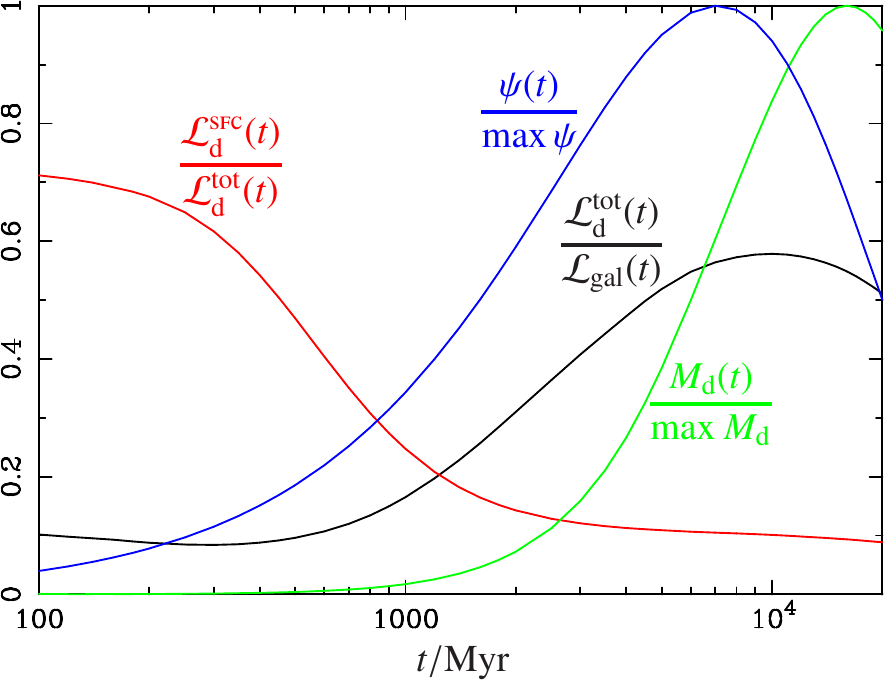}
  \caption{\label{fig:Dust_Evolution}%
    Evolution, as a function of galactic age~$t$, of the following
    quantities for the Milky Way model:
    star formation rate $\psi(t)$ (relative to its maximal value);
    dust mass $M_\dust(t)$ (idem);
    ratio of the bolometric luminosity emitted by dust in the whole galaxy,
    $\Lbol_\dust^\total(t)$, to the overall bolometric luminosity of the
    galaxy, $\Lbol_\galaxy(t)$;
    ratio of the bolometric luminosity emitted by dust in star-forming clouds,
    $\Lbol_\dust^\SFC(t)$, to $\Lbol_\dust^\total(t)$.
  }
\end{figure}

In addition to the overall SED of a galaxy,
the code may on request separately output the SEDs of its several components
\mbox{--~}stars, ionized gas, dust species~\mbox{--,}
whether in the diffuse medium or in star-forming regions.
This is illustrated by Fig.~\ref{fig:Composantes_13Gyr} for the Milky Way
model at $t = 13\mult\Gyr$.
At this late age, UV-to-\mbox{near-IR} wavelengths are dominated by the
attenuated light of stars, with a negligible contribution from the nebular
continuum, and \mbox{mid-IR}-to-submm wavelengths are dominated by dust grains.
Whereas the emission of grains in star-forming clouds is of the same order
of magnitude in the \mbox{mid-IR} as that of grains in the diffuse ISM, the
contribution of the latter is overwhelming in the \mbox{far-IR}\slash submm
domain and in the overall emission of dust, as
may be seen in Fig.~\ref{fig:Dust_Evolution}.

\begin{figure}
  \includegraphics[width=\columnwidth]{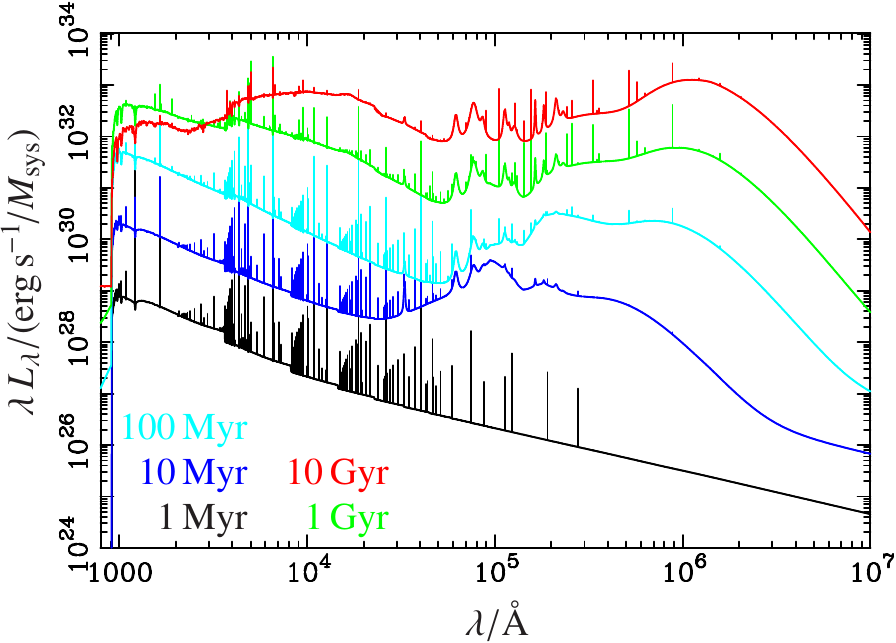}
  \caption{\label{fig:Spiral_template}%
    Spectral energy distribution of the Milky Way model at various ages
    (color coded on the plot).
  }
\end{figure}

The evolution of the Milky Way model's SED is plotted in
Fig.~\ref{fig:Spiral_template} from the \mbox{far-UV} to submm~wavelengths.
We analyze it below with the help of Fig.~\ref{fig:Dust_Evolution}, where
we have plotted, for the same scenario, the evolution as a function of age
of the following quantities:
the ratio of the bolometric luminosity radiated by dust grains in all regions
of the galaxy, $\Lbol_\dust^\total(t)$, to the bolometric luminosity
of the whole galaxy (stars, gas and dust), $\Lbol_\galaxy(t)$;
the ratio of the bolometric luminosity radiated by grains in star-forming
clouds, $\Lbol_\dust^\SFC(t)$, to $\Lbol_\dust^\total(t)$;
the dust mass $M_\dust(t)$;
the star formation rate $\psi(t)$.
(The last two quantities are normalized to their maximal values.)

Three trends appearing in Fig.~\ref{fig:Spiral_template} deserve to be
especially emphasized:
Firstly, the luminosity increases in the UV and, even more, in the
optical--\mbox{near-IR} domain until late ages.
This happens because the modeled galaxy assembles by infall on a long
timescale:
the star formation rate (SFR) being regulated by the mass of gas available
in the ISM, it initially increases during several~$\Gyr$
(see Fig.~\ref{fig:Dust_Evolution}) and then decreases only slowly;
as a result, in this model, the rapid decline of the luminosity of a
single stellar population is more than balanced by the accumulation of
successive stellar generations, and the overall luminosity grows steadily
at wavelengths dominated by stars.

Secondly, the spectrum progressively reddens from the UV to the near-infrared,
due to the aging of the bulk of the stellar population and to the increasing
metallicity of the ISM from which stars form.
In parallel, the discontinuities of the nebular continuum are more and more
swamped in the stellar emission, up to the point, after $1\mult\Gyr$, where
they are not discernible anymore.

Lastly, at longer wavelengths, the most noticeable feature is the shift of the
infrared peak%
\footnote{%
  In the \PEGASE.3 scenarios fitted to high-redshift radiogalaxy hosts by
  \citet{RoccaVolmerange2013}, which use
  $\mathord{\sim}\,1\mult\Gyr$~timescales for star formation and infall,
  the \mbox{far-IR} peak is much more intense at young ages and highly
  sensitive to outflow episodes.
}
from \mbox{$\lambda \simeq 10\mult \micron$} at an age of $10\mult\Myr$
to more than $100\mult \micron$ at $10\mult\Gyr$.
(Note that this peak is distinct from the near-IR peak observed
around $\lambda \in [1, 2]\mult \micron$ at late ages
(see Fig.~\ref{fig:Spiral_template}): the latter
is caused by evolved low-mass cold giant stars.)
At young ages, most of the emission by dust comes from grains
in star-forming regions, heated to very high temperatures by the UV
photons produced by nearby young luminous stars, and reradiating
in the \mbox{mid-IR}.
After $1\mult\Gyr$, most of the dust emission is due to grains in the diffuse
ISM and radiating in the \mbox{far-IR} (see the red curve in
Fig.~\ref{fig:Dust_Evolution}).

The reasons for this behavior are the following:
As time progresses, the total mass in the old and intermediate-age stars
dominating optical wavelengths and scattered through the whole galaxy grows.
The radiation field produced by these stars becomes therefore more intense.
It is however softer than in star-forming regions
(see Fig.~\ref{fig:radiation_field}), so grains in the diffuse medium
reach colder temperatures and emit at longer wavelengths.
In parallel, the ISM is enriched by the ejecta of previous stellar
generations, and its metallicity constantly increases.
The mass of the ISM grows until $\mathord{\approx}\,7\mult\Gyr$
(see the blue curve in Fig.~\ref{fig:Dust_Evolution}:
we remind the reader 
that, for the star formation law adopted in the Milky Way model,
the ISM mass is proportional to the SFR $\psi(t)$).
The mass of metals in the ISM, which is equal to the product of the metallicity
of the ISM by the mass of the latter, peaks therefore later than the SFR.
The same holds consequently for the mass of dust (green curve in
Fig.~\ref{fig:Dust_Evolution}).
The ratio of $\Lbol_\dust^\total(t)$ to $\Lbol_\galaxy(t)$
(black curve) \mbox{--~}which depends on
recent star formation, on the mass in older stars and on that of dust~--
reaches its maximum at an age between the peak ages of the SFR
and of the mass of dust.
\subsection{Temperatures and SEDs of grains}
\begin{figure}
  \includegraphics[width=\columnwidth]{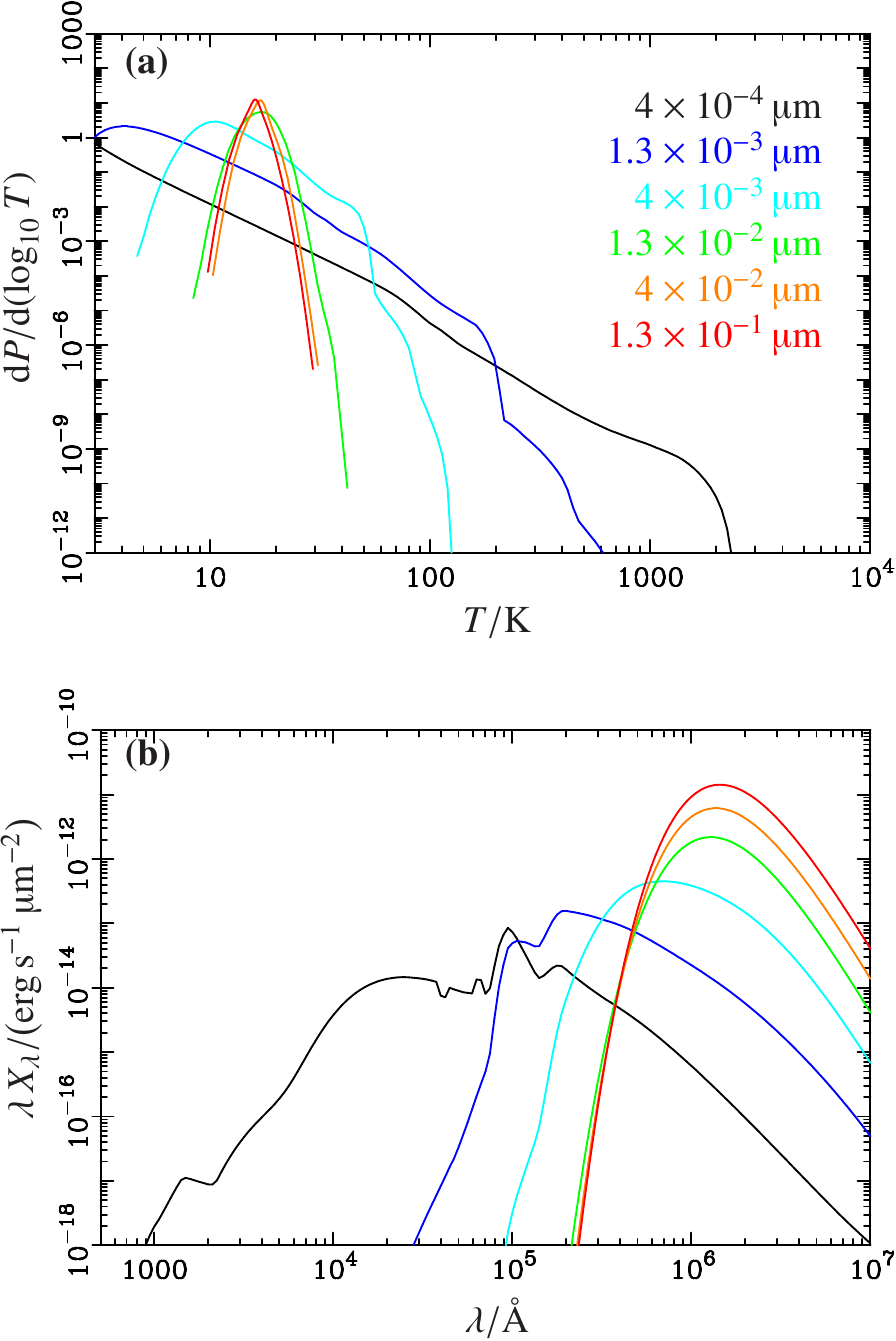}
  \caption{\label{fig:GrainSpectra}%
    \textbf{(a)}~Temperature probability distribution of silicate grains
    in the diffuse medium of the $13\mult\Gyr$-old Milky Way model.
    Grain radii are color coded as indicated on the plot.
    \endgraf
    \textbf{(b)} Emission spectra of these grains (same color coding for grain
    radii as in previous subfigure).
    The quantity $X_\lambda$ is the spectral exitance (luminosity per unit surface of the grain).}
\end{figure}

\PEGASE.3 may optionally provide the temperature probability distribution
of stochastically heated individual grains and their emission spectrum.
As an illustration, Fig.~\ref{fig:GrainSpectra} shows these properties
for silicate grains with various radii in the diffuse ISM of the Milky Way
model at $13\mult\Gyr$.
This figure highlights that small grains have a large range of temperatures
and emit significantly at short wavelengths.
For bigger and bigger grains, the temperature distribution narrows around
the equilibrium value, and the emission spectrum becomes similar to that
of a blackbody peaking in the \mbox{far-IR}.

\begin{figure}
  \includegraphics[width=\columnwidth]{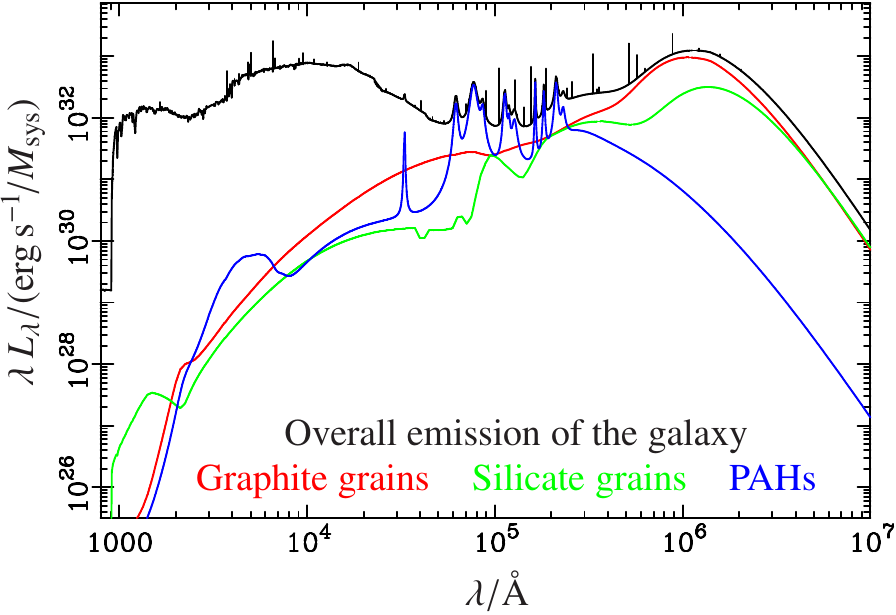}
  \caption{\label{fig:dustgrains}%
    Spectral energy distributions of the several species of dust grains
    (graphites, red line; silicates, green line; PAHs, blue line)
    and of the whole galaxy (black line) for the $13\mult\Gyr$-old Milky Way
    model.}
\end{figure}

The SEDs of the several species of dust grains \mbox{--~}silicates, graphites
and PAHs~-- are plotted in Fig.~\ref{fig:dustgrains} for the same model.
While PAH features are prominent in the \mbox{mid-IR}, graphites dominate
at longer wavelengths.
Only in the submillimeter domain are silicates major contributors to the
global SED.
\section{Conclusion and prospects%
  \label{sec:conclusion}}
This paper presents \PEGASE.3, a new version of the code \PEGASE\
specifically aimed to model the spectrochemical evolution of galaxies
at all redshifts.
The most important outputs of the code are synthetic SEDs from the \mbox{far-UV}
to the submillimeter, from which colors may be determined in a large number
of photometric systems.
Because the star formation history and the chemical evolution are computed
consistently, this large wavelength range should help to lessen the degeneracy
between the parameters
(e.g., stellar mass, current SFR, age, metallicity)
derived from fits of model SEDs to galaxy observations.

The main improvement, with respect to previous versions of the code,
is
the extension to the mid-~and far-infrared, which required to model the
evolution of the dust content and its effects on the light radiated by a galaxy.
The computation of the nebular emission, in particular of metallic
and infrared lines,
has also been
entirely upgraded using \textsc{Cloudy}.

To determine the amount of carbonaceous and silicate grains,
\PEGASE\ now follows the detailed evolution of the abundances of the
most important elements in the interstellar medium.
To this purpose, two models are proposed in the code:
a phenomenological one in which the mass of grains is directly related
to the amount of their constituents in the ISM;
a more physical one, based on \citet{Dwek1998} and fitted by him to Milky Way
data, in which dust grains form in the circumstellar envelopes of stars and
around supernovae, accrete on grains already present in the ISM and are
destroyed by supernovae blast waves in the ISM.

The overall optical properties of dust are then computed,
assuming some size distributions for the various species of grains.
These optical properties are used to obtain the attenuation through
the diffuse interstellar medium of the stellar and nebular light at all
wavelengths, using grids of radiative transfer for geometries appropriate
to either spheroidal galaxies or spiral disks and bulges.
All these grids were computed beforehand with Monte Carlo simulations based
on the method of virtual interactions.
The attenuation by dust in their birth cloud of the light emitted by young
stars, in particular in the Lyman continuum, is also estimated.

The advantage of this procedure is that it also provides the mean radiation
field in the diffuse medium and the one averaged over all star-forming clouds.
The re-emission by dust of the energy it absorbed is then computed from
the optical properties of individual grains, taking into account their
stochastic heating by the mean radiation field in the two regions.

In spite of some limitations (use of mean radiation fields, self-absorption
by dust grains not rigorously treated), the method implemented in \PEGASE\
to compute \mbox{far-UV}-to-submm SEDs is more physical than in other codes,
where the shape of the radiation field does not evolve
\citep[e.g.,][]{Galliano2011},
or which use template attenuation curves unrelated to the dust content
(for instance, the one determined for starbursts by \citealp{Calzetti1994})
or template infrared SEDs (e.g., graybodies, as in \citealp{DaCunha2008},
or the semi-empirical SEDs of \citealp{Dale2014}).
It is beyond the scope of this paper to compare \PEGASE.3 to all these codes,
but, to highlight the specifities of ours, let us just consider a recent one,
\textsc{Cigale} \citep[in its last version]{Cigale2018}.
The working principle of this code is quite different from that of
\PEGASE.3, despite similar goals, as, when fitting modeled SEDs
to an object's observed spectral data, \textsc{Cigale} employs a powerful
Bayesian method to derive the likelihood of the input parameters.
However, in this procedure, the various galactic components
\mbox{--~}stars, gas and dust~-- are treated independently.
For instance, the metallicity of stellar populations is constant,
and the star formation history is quite arbitrary since neither the evolution
of the ISM nor the history of mass assembly of the galaxy are considered.
The only theoretical constraint is that the energy balance between the
absorption of stellar and nebular light by dust and its re-emission
must be respected.
This data-driven approach provides a lot of suppleness
but does not ensure the internal consistency of the models.
On the other hand, although \PEGASE.3 does not incorporate a fitting
procedure of the code's scenarios to observed data%
\footnote{However, the code \textsc{Z-P\'eg} of \citet{ZPEG} (available at
  \ifdefined\href
  \href{www.iap.fr/pegase/}{\texttt{www.iap.fr\slash pegase/}}%
  \else
  \texttt{www.iap.fr\slash pegase/}\fi),
  which currently uses \PEGASE.2 templates
  from \citet{Pegase.2} to estimate by \mbox{$\chi^2$-minimization} both
  the photometric redshift of an object and the best-fitting model,
  might easily be updated to process \PEGASE.3 outputs.},
it strives to model simultaneously the evolution of stellar populations
and of gas and dust in the ISM, in particular the abundances of various
metals and types of grains, as well as the contributions of these components
to the SED.
Finally, \textsc{Cigale} relies on libraries of single stellar populations
computed by either \citet{BC2003} or \citet{Maraston2005};
in contrast to \PEGASE, users of this code do not,
therefore, have full control of these inputs and
are restricted to a limited number of initial mass functions.

Codes more sophisticated than \PEGASE\ in some respects also exist.
For instance, \textsc{Grasil} \citep{Granato2000}
takes into account the radiation field at each point in the galaxy
and the self-absorption by dust grains.
This, however, requires users to run anew Monte Carlo simulations
of radiative transfer at all ages, for each modeled galaxy,
which seems impractical to simulate the evolution of a large number of
objects on cosmological timescales.
To fulfill this aim, the choice made in \PEGASE.3 was
rather to try to strike
the right balance between computational efficiency and physical correctness,
while still maintaining consistency of the modeling.

The code provides a wealth of outputs, besides SEDs and derived quantities
(colors, equivalent widths of emission lines, etc.):
among others, the masses in stars, compact stellar remnants, the ISM and
dust grains;
the abundances of the most important elements;
the rates of star formation, ionizing photon emission and supernova explosions,
both for core-collapse and type~Ia objects.
Because of the large amount of space they would occupy, other outputs are
only optional, such as the separate SEDs emitted by dust species,
the mean radiation fields in the diffuse medium and in star-forming clouds,
or, even more, the temperature distributions and SEDs of individual grains.
Some of these outputs were illustrated in this paper for the ``Milky Way''
model.

Preliminary versions of \PEGASE.3 were also used to study spheroidal galaxies.
In particular, \citet{RoccaVolmerange2013} analyzed the SEDs of radiogalaxy
hosts observed by \emph{HST}, \emph{Spitzer} and \emph{Herschel}
at redshifts $z \in [1, 5]$;
the authors showed that these distant objects were already old then
(\mbox{$\text{age}\geqslant 1\mult\Gyr$}),
massive (stellar mass~$\mathord{\sim}\,10^{11}$ to $10^{12}\mult M_\odot$)
and that they had recently undergone an extended burst of star formation.
\citet{Drouart2016} and \citet{Podigachoski2016} also used \PEGASE.3,
in combination with models of AGNs, to disentangle the contributions
of an active nucleus and of dust heated by young stars to the mid-~and
\mbox{far-IR} emission.

Because of its large spectral coverage, the code provides a nearly
``bolometrically''-complete modeling for standard galaxies.
This will make it possible
to study in a more consistent way the star formation history,
the dust attenuation and emission and the chemical evolution in these objects,
and to build more-reliable indicators of the recent star formation rate,
stellar mass, age, metallicity and dust content.

The flexibility of the code and the variety of the scenarios that may be
input should be of particular interest for cosmological simulations.
For instance, the stellar initial mass function and the parameters of the
laws giving the star formation, infall and outflow rates may easily be changed.
Users may also provide a file giving these rates at some ages, from
which the code interpolates at intermediate ages.
They may even define several episodes of star formation, infall and outflow,
whether consecutive or overlapping, and modulate the star formation
rate stochastically.

The current spectral resolution of model SEDs is low but will be improved
in the near future by our team.
We also intend to implement in \PEGASE\ more-realistic models of dust grain
composition and evolution, such as the one built with \textsc{Themis}
\citep{THEMIS}, and more-modern sets of stellar evolutionary tracks,
spectra and chemical yields.
However, in view of analyzing forthcoming infrared data, for example those that
the \emph{SPICA} space observatory will provide \citep{SPICA} if selected
for launch, our main effort will concentrate on the modeling of star-forming
regions, in particular the competition between gas and dust in \HII~regions to
absorb ionizing photons and the effects of grains on nebular emission.
\begin{acknowledgements}
  Michel Fioc is grateful to the NASA/Goddard Space Flight Center
  (Greenbelt, Maryland) for its hospitality during the early phases
  of this work.
  In particular, he warmly thanks Eli Dwek, his supervisor during this stay,
  for his help in the modeling of dust grains and, much more
  importantly, for his kindness.

  Brigitte Rocca-Volmerange acknowledges financial support from the CNES-PRSS
  program to her works on the analysis with \PEGASE.3 of spatial observations
  of distant radiogalaxies.
  The authors also thank the referee for insightful questions on the modeling
  of star-forming regions and nebular emission:
  they helped to clarify the paper and incited them to significantly improve
  the code.
\end{acknowledgements}
\bibliographystyle{aa}
\bibliography{references,references2,references3}

\begin{thebibliography}{129}
\expandafter\ifx\csname natexlab\endcsname\relax\def\natexlab#1{#1}\fi

\bibitem[{{Alloin} {et~al.}(1971){Alloin}, {Andrillat}, \&
  {Souffrin}}]{Alloin+1971}
{Alloin}, D., {Andrillat}, Y., \& {Souffrin}, S. 1971, \aap, 10, 401

\bibitem[{{Althaus} \& {Benvenuto}(1997)}]{Althaus1997}
{Althaus}, L.~G. \& {Benvenuto}, O.~G. 1997, \apj, 477, 313

\bibitem[{{Anders} \& {Grevesse}(1989)}]{Anders1989}
{Anders}, E. \& {Grevesse}, N. 1989, \gca, 53, 197

\bibitem[{{Battisti} {et~al.}(2016){Battisti}, {Calzetti}, \&
  {Chary}}]{Battisti2016}
{Battisti}, A.~J., {Calzetti}, D., \& {Chary}, R.-R. 2016, \apj, 818, 13

\bibitem[{{Binney} \& {Tremaine}(2008)}]{Binney2008}
{Binney}, J. \& {Tremaine}, S. 2008, {Galactic Dynamics: Second Edition}
  (Princeton University Press)

\bibitem[{{Bl\"ocker}(1995)}]{Bloecker1995}
{Bl\"ocker}, T. 1995, \aap, 299, 755

\bibitem[{{Boissier} \& {Prantzos}(1999)}]{Boissier}
{Boissier}, S. \& {Prantzos}, N. 1999, \mnras, 307, 857

\bibitem[{{Boissier} \& {Prantzos}(2000)}]{Boissier+Prantzos}
{Boissier}, S. \& {Prantzos}, N. 2000, \mnras, 312, 398

\bibitem[{{Boquien} {et~al.}(2019){Boquien}, {Burgarella}, {Roehlly}, {Buat},
  {Ciesla}, {Corre}, {Inoue}, \& {Salas}}]{Cigale2018}
{Boquien}, M., {Burgarella}, D., {Roehlly}, Y., {et~al.} 2019, \aap, 622, A103

\bibitem[{{Bressan} {et~al.}(1993){Bressan}, {Fagotto}, {Bertelli}, \&
  {Chiosi}}]{Padova0.02}
{Bressan}, A., {Fagotto}, F., {Bertelli}, G., \& {Chiosi}, C. 1993, \aaps, 100,
  647

\bibitem[{{Bressan} {et~al.}(2012){Bressan}, {Marigo}, {Girardi}, {Salasnich},
  {Dal Cero}, {Rubele}, \& {Nanni}}]{PARSEC}
{Bressan}, A., {Marigo}, P., {Girardi}, L., {et~al.} 2012, \mnras, 427, 127

\bibitem[{{Bruzual}(1983)}]{Bruzual1983b}
{Bruzual}, G. 1983, \apj, 273, 105

\bibitem[{{Bruzual} \& {Charlot}(2003)}]{BC2003}
{Bruzual}, G. \& {Charlot}, S. 2003, \mnras, 344, 1000

\bibitem[{{Buat} {et~al.}(2014){Buat}, {Heinis}, {Boquien}, {Burgarella},
  {Charmandaris}, {Boissier}, {Boselli}, {Le Borgne}, \& {Morrison}}]{Buat2014}
{Buat}, V., {Heinis}, S., {Boquien}, M., {et~al.} 2014, \aap, 561, A39

\bibitem[{{Buta} \& {Williams}(1995)}]{Buta1995}
{Buta}, R. \& {Williams}, K.~L. 1995, \aj, 109, 543

\bibitem[{{Calzetti} {et~al.}(1994){Calzetti}, {Kinney}, \&
  {Storchi-Bergmann}}]{Calzetti1994}
{Calzetti}, D., {Kinney}, A.~L., \& {Storchi-Bergmann}, T. 1994, \apj, 429, 582

\bibitem[{{Cassar{\`a}} {et~al.}(2015){Cassar{\`a}}, {Piovan}, \&
  {Chiosi}}]{Cassara2015}
{Cassar{\`a}}, L.~P., {Piovan}, L., \& {Chiosi}, C. 2015, \mnras, 450, 2231

\bibitem[{{Chabrier} \& {Baraffe}(1997)}]{Chabrier1997}
{Chabrier}, G. \& {Baraffe}, I. 1997, \aap, 327, 1039

\bibitem[{{Charlot} \& {Bruzual}(1991)}]{CB1991}
{Charlot}, S. \& {Bruzual}, G. 1991, \apj, 367, 126

\bibitem[{{Charlot} \& {Fall}(2000)}]{CharlotFall2000}
{Charlot}, S. \& {Fall}, S.~M. 2000, \apj, 539, 718

\bibitem[{{Charlot} \& {Longhetti}(2001)}]{Charlot+Longhetti2001}
{Charlot}, S. \& {Longhetti}, M. 2001, \mnras, 323, 887

\bibitem[{{Chevallard} \& {Charlot}(2016)}]{Chevallard+Charlot2016}
{Chevallard}, J. \& {Charlot}, S. 2016, \mnras, 462, 1415

\bibitem[{{Ciesla} {et~al.}(2016){Ciesla}, {Boselli}, {Elbaz}, {Boissier},
  {Buat}, {Charmandaris}, {Schreiber}, {B{\'e}thermin}, {Baes}, {Boquien}, {De
  Looze}, {Fern{\'a}ndez-Ontiveros}, {Pappalardo}, {Spinoglio}, \&
  {Viaene}}]{Ciesla+2016}
{Ciesla}, L., {Boselli}, A., {Elbaz}, D., {et~al.} 2016, \aap, 585, A43

\bibitem[{{Conroy}(2010)}]{Conroy2010}
{Conroy}, C. 2010, \mnras, 404, 247

\bibitem[{{Conroy}(2013)}]{Conroy_ARAA}
{Conroy}, C. 2013, \araa, 51, 393

\bibitem[{{Cousin} {et~al.}(2015){Cousin}, {Lagache}, {Bethermin}, {Blaizot},
  \& {Guiderdoni}}]{Cousin+2015}
{Cousin}, M., {Lagache}, G., {Bethermin}, M., {Blaizot}, J., \& {Guiderdoni},
  B. 2015, \aap, 575, A32

\bibitem[{{da Cunha} {et~al.}(2008){da Cunha}, {Charlot}, \&
  {Elbaz}}]{DaCunha2008}
{da Cunha}, E., {Charlot}, S., \& {Elbaz}, D. 2008, \mnras, 388, 1595

\bibitem[{{Dale} {et~al.}(2014){Dale}, {Helou}, {Magdis}, {Armus},
  {D{\'{\i}}az-Santos}, \& {Shi}}]{Dale2014}
{Dale}, D.~A., {Helou}, G., {Magdis}, G.~E., {et~al.} 2014, \apj, 784, 83

\bibitem[{{Davidson}(1977)}]{Davidson77}
{Davidson}, K. 1977, \apj, 218, 20

\bibitem[{{de Vaucouleurs}(1991)}]{Vaucouleurs1991}
{de Vaucouleurs}, G. 1991, Science, 254, 1667

\bibitem[{{de Vaucouleurs} \& {Pence}(1978)}]{deVaucouleurs+Pence}
{de Vaucouleurs}, G. \& {Pence}, W.~D. 1978, \aj, 83, 1163

\bibitem[{{Devriendt} {et~al.}(1999){Devriendt}, {Guiderdoni}, \&
  {Sadat}}]{Devriendt1999}
{Devriendt}, J.~E.~G., {Guiderdoni}, B., \& {Sadat}, R. 1999, \aap, 350, 381

\bibitem[{{Dopita} {et~al.}(2006){Dopita}, {Fischera}, {Sutherland}, {Kewley},
  {Tuffs}, {Popescu}, {van Breugel}, {Groves}, \& {Leitherer}}]{Dopita+2006}
{Dopita}, M.~A., {Fischera}, J., {Sutherland}, R.~S., {et~al.} 2006, \apj, 647,
  244

\bibitem[{{Dopita} {et~al.}(2005){Dopita}, {Groves}, {Fischera}, {Sutherland},
  {Tuffs}, {Popescu}, {Kewley}, {Reuland}, \& {Leitherer}}]{Dopita2005b}
{Dopita}, M.~A., {Groves}, B.~A., {Fischera}, J., {et~al.} 2005, \apj, 619, 755

\bibitem[{{Draine} \& {Lee}(1984)}]{Draine1984}
{Draine}, B.~T. \& {Lee}, H.~M. 1984, \apj, 285, 89

\bibitem[{{Draine} \& {Li}(2001)}]{Draine2001}
{Draine}, B.~T. \& {Li}, A. 2001, \apj, 551, 807

\bibitem[{{Drouart} {et~al.}(2016){Drouart}, {Rocca-Volmerange}, {De Breuck},
  {Fioc}, {Lehnert}, {Seymour}, {Stern}, \& {Vernet}}]{Drouart2016}
{Drouart}, G., {Rocca-Volmerange}, B., {De Breuck}, C., {et~al.} 2016, \aap,
  593, A109

\bibitem[{{Dwek}(1998)}]{Dwek1998}
{Dwek}, E. 1998, \apj, 501, 643

\bibitem[{{Eldridge} {et~al.}(2008){Eldridge}, {Izzard}, \& {Tout}}]{Eldridge}
{Eldridge}, J.~J., {Izzard}, R.~G., \& {Tout}, C.~A. 2008, \mnras, 384, 1109

\bibitem[{{Eldridge} {et~al.}(2017){Eldridge}, {Stanway}, {Xiao}, {McClelland},
  {Taylor}, {Ng}, {Greis}, \& {Bray}}]{Eldridge+2017}
{Eldridge}, J.~J., {Stanway}, E.~R., {Xiao}, L., {et~al.} 2017, \pasa, 34, e058

\bibitem[{{Fagotto} {et~al.}(1994{\natexlab{a}}){Fagotto}, {Bressan},
  {Bertelli}, \& {Chiosi}}]{Padova0.0004-0.05}
{Fagotto}, F., {Bressan}, A., {Bertelli}, G., \& {Chiosi}, C.
  1994{\natexlab{a}}, \aaps, 104, 365

\bibitem[{{Fagotto} {et~al.}(1994{\natexlab{b}}){Fagotto}, {Bressan},
  {Bertelli}, \& {Chiosi}}]{Padova0.004-0.008}
{Fagotto}, F., {Bressan}, A., {Bertelli}, G., \& {Chiosi}, C.
  1994{\natexlab{b}}, \aaps, 105, 29

\bibitem[{{Fagotto} {et~al.}(1994{\natexlab{c}}){Fagotto}, {Bressan},
  {Bertelli}, \& {Chiosi}}]{Padova0.1}
{Fagotto}, F., {Bressan}, A., {Bertelli}, G., \& {Chiosi}, C.
  1994{\natexlab{c}}, \aaps, 105, 39

\bibitem[{{Ferland} {et~al.}(2017){Ferland}, {Chatzikos}, {Guzm{\'a}n},
  {Lykins}, {van Hoof}, {Williams}, {Abel}, {Badnell}, {Keenan}, {Porter}, \&
  {Stancil}}]{Cloudy}
{Ferland}, G.~J., {Chatzikos}, M., {Guzm{\'a}n}, F., {et~al.} 2017, \rmxaa, 53,
  385

\bibitem[{{Fern{\'a}ndez-Ontiveros} {et~al.}(2017){Fern{\'a}ndez-Ontiveros},
  {Armus}, {Baes}, {Bernard-Salas}, {Bolatto}, {Braine}, {Ciesla}, {De Looze},
  {Egami}, {Fischer}, {Giard}, {Gonz{\'a}lez-Alfonso}, {Granato}, {Gruppioni},
  {Imanishi}, {Ishihara}, {Kaneda}, {Madden}, {Malkan}, {Matsuhara},
  {Matsuura}, {Nagao}, {Najarro}, {Nakagawa}, {Onaka}, {Oyabu},
  {Pereira-Santaella}, {P{\'e}rez Fournon}, {Roelfsema}, {Santini}, {Silva},
  {Smith}, {Spinoglio}, {van der Tak}, {Wada}, \& {Wu}}]{SPICA}
{Fern{\'a}ndez-Ontiveros}, J.~A., {Armus}, L., {Baes}, M., {et~al.} 2017,
  \pasa, 34, e053

\bibitem[{{Fioc}(1997)}]{these_Fioc}
{Fioc}, M. 1997, {Ph.D. thesis}, Universit\'e Paris~XI, in French. Available at
  \ifdefined\href
  \href{ftp://ftp.iap.fr/pub/from\string_users/fioc/these.pdf}{%
  \texttt{ftp://ftp.iap.fr\slash pub\slash from\string_users\slash fioc\slash
  these.pdf}}\else \texttt{ftp://ftp.iap.fr\slash pub\slash
  from\string_users\slash fioc\slash these.pdf}\fi

\bibitem[{{Fioc} \& {Rocca-Volmerange}(1997)}]{Pegase.1}
{Fioc}, M. \& {Rocca-Volmerange}, B. 1997, \aap, 326, 950

\bibitem[{{Fioc} \& {Rocca-Volmerange}(1999{\natexlab{a}})}]{Fioc1999b}
{Fioc}, M. \& {Rocca-Volmerange}, B. 1999{\natexlab{a}}, \aap, 351, 869

\bibitem[{{Fioc} \& {Rocca-Volmerange}(1999{\natexlab{b}})}]{Pegase.2}
{Fioc}, M. \& {Rocca-Volmerange}, B. 1999{\natexlab{b}}, arXiv:astro-ph/9912179

\bibitem[{{Fioc} \& {Rocca-Volmerange}(2019)}]{doc_Pegase.3}
{Fioc}, M. \& {Rocca-Volmerange}, B. 2019, {The P\'egase.3 code of
  spectrochemical evolution of galaxies: documentation and complements},
  \ifdefined\href
  \href{https://arxiv.org/abs/1902.02198}{arXiv:1902.02198}\else
  arXiv:1902.02198\fi

\bibitem[{{Flores-Fajardo} {et~al.}(2011){Flores-Fajardo}, {Morisset},
  {Stasi{\'n}ska}, \& {Binette}}]{U_DISM}
{Flores-Fajardo}, N., {Morisset}, C., {Stasi{\'n}ska}, G., \& {Binette}, L.
  2011, \mnras, 415, 2182

\bibitem[{{Fr\"ohlich}(1982)}]{Froehlich}
{Fr\"ohlich}, H.-E. 1982, Astronomische Nachrichten, 303, 97

\bibitem[{{Galliano} {et~al.}(2008){Galliano}, {Dwek}, \&
  {Chanial}}]{Galliano2008}
{Galliano}, F., {Dwek}, E., \& {Chanial}, P. 2008, \apj, 672, 214

\bibitem[{{Galliano} {et~al.}(2011){Galliano}, {Hony}, {Bernard}, {Bot},
  {Madden}, {Roman-Duval}, {Galametz}, {Li}, {Meixner}, {Engelbracht},
  {Lebouteiller}, {Misselt}, {Montiel}, {Panuzzo}, {Reach}, \&
  {Skibba}}]{Galliano2011}
{Galliano}, F., {Hony}, S., {Bernard}, J.-P., {et~al.} 2011, \aap, 536, A88

\bibitem[{{Girardi} {et~al.}(1996){Girardi}, {Bressan}, {Chiosi}, {Bertelli},
  \& {Nasi}}]{Padova0.0001}
{Girardi}, L., {Bressan}, A., {Chiosi}, C., {Bertelli}, G., \& {Nasi}, E. 1996,
  \aaps, 117, 113

\bibitem[{{Graham} \& {Worley}(2008)}]{Graham2008}
{Graham}, A.~W. \& {Worley}, C.~C. 2008, \mnras, 388, 1708

\bibitem[{{Granato} {et~al.}(2000){Granato}, {Lacey}, {Silva}, {Bressan},
  {Baugh}, {Cole}, \& {Frenk}}]{Granato2000}
{Granato}, G.~L., {Lacey}, C.~G., {Silva}, L., {et~al.} 2000, \apj, 542, 710

\bibitem[{{Greggio} \& {Renzini}(1983)}]{Greggio+Renzini}
{Greggio}, L. \& {Renzini}, A. 1983, \aap, 118, 217

\bibitem[{{Groenewegen} \& {de Jong}(1993)}]{Groenewegen1993}
{Groenewegen}, M.~A.~T. \& {de Jong}, T. 1993, \aap, 267, 410

\bibitem[{{Guhathakurta} \& {Draine}(1989)}]{Guhathakurta1989}
{Guhathakurta}, P. \& {Draine}, B.~T. 1989, \apj, 345, 230

\bibitem[{{Guiderdoni} \& {Rocca-Volmerange}(1987)}]{GRV1987}
{Guiderdoni}, B. \& {Rocca-Volmerange}, B. 1987, \aap, 186, 1

\bibitem[{{Gutkin} {et~al.}(2016){Gutkin}, {Charlot}, \&
  {Bruzual}}]{Gutkin+2016}
{Gutkin}, J., {Charlot}, S., \& {Bruzual}, G. 2016, \mnras, 462, 1757

\bibitem[{{Hatton} {et~al.}(2003){Hatton}, {Devriendt}, {Ninin}, {Bouchet},
  {Guiderdoni}, \& {Vibert}}]{Hatton+2003}
{Hatton}, S., {Devriendt}, J.~E.~G., {Ninin}, S., {et~al.} 2003, \mnras, 343,
  75

\bibitem[{{Henry} \& {Worthey}(1999)}]{HW}
{Henry}, R.~B.~C. \& {Worthey}, G. 1999, \pasp, 111, 919

\bibitem[{{Henyey} \& {Greenstein}(1941)}]{HG}
{Henyey}, L.~G. \& {Greenstein}, J.~L. 1941, \apj, 93, 70

\bibitem[{{Hodge}(1983)}]{Hodge83}
{Hodge}, P.~W. 1983, \pasp, 95, 721

\bibitem[{{Jones} {et~al.}(2017){Jones}, {K{\"o}hler}, {Ysard}, {Bocchio}, \&
  {Verstraete}}]{THEMIS}
{Jones}, A.~P., {K{\"o}hler}, M., {Ysard}, N., {Bocchio}, M., \& {Verstraete},
  L. 2017, \aap, 602, A46

\bibitem[{{Karakas}(2010)}]{Karakas2010}
{Karakas}, A.~I. 2010, \mnras, 403, 1413

\bibitem[{{Koester} \& {Sch\"onberner}(1986)}]{Koester1986}
{Koester}, D. \& {Sch\"onberner}, D. 1986, \aap, 154, 125

\bibitem[{{Kroupa} {et~al.}(1993){Kroupa}, {Tout}, \& {Gilmore}}]{Kroupa1993}
{Kroupa}, P., {Tout}, C.~A., \& {Gilmore}, G. 1993, \mnras, 262, 545

\bibitem[{{Kurucz}(1979)}]{Kurucz1979}
{Kurucz}, R.~L. 1979, \apjs, 40, 1

\bibitem[{{Lacey} {et~al.}(2008){Lacey}, {Baugh}, {Frenk}, {Silva}, {Granato},
  \& {Bressan}}]{Lacey2008}
{Lacey}, C.~G., {Baugh}, C.~M., {Frenk}, C.~S., {et~al.} 2008, \mnras, 385,
  1155

\bibitem[{{Laor} \& {Draine}(1993)}]{Laor1993}
{Laor}, A. \& {Draine}, B.~T. 1993, \apj, 402, 441

\bibitem[{{Larson}(1972)}]{Larson72}
{Larson}, R.~B. 1972, Nature Physical Science, 236, 7

\bibitem[{{Le Borgne} \& {Rocca-Volmerange}(2002)}]{ZPEG}
{Le Borgne}, D. \& {Rocca-Volmerange}, B. 2002, \aap, 386, 446

\bibitem[{{Le Borgne} {et~al.}(2004){Le Borgne}, {Rocca-Volmerange},
  {Prugniel}, {Lan{\c c}on}, {Fioc}, \& {Soubiran}}]{LeBorgne2004}
{Le Borgne}, D., {Rocca-Volmerange}, B., {Prugniel}, P., {et~al.} 2004, \aap,
  425, 881

\bibitem[{{L\'eger} \& {Puget}(1984)}]{LegerPuget1984}
{L\'eger}, A. \& {Puget}, J.~L. 1984, \aap, 137, L5

\bibitem[{{Leitherer} {et~al.}(2014){Leitherer}, {Ekstr{\"o}m}, {Meynet},
  {Schaerer}, {Agienko}, \& {Levesque}}]{Leitherer2014}
{Leitherer}, C., {Ekstr{\"o}m}, S., {Meynet}, G., {et~al.} 2014, \apjs, 212, 14

\bibitem[{{Leitherer} {et~al.}(1999){Leitherer}, {Schaerer}, {Goldader},
  {Delgado}, {Robert}, {Kune}, {de Mello}, {Devost}, \&
  {Heckman}}]{Leitherer1999}
{Leitherer}, C., {Schaerer}, D., {Goldader}, J.~D., {et~al.} 1999, \apjs, 123,
  3

\bibitem[{{Lejeune} {et~al.}(1998){Lejeune}, {Cuisinier}, \&
  {Buser}}]{Lejeune1998}
{Lejeune}, T., {Cuisinier}, F., \& {Buser}, R. 1998, \aaps, 130, 65

\bibitem[{{Li} \& {Draine}(2001)}]{Li2001}
{Li}, A. \& {Draine}, B.~T. 2001, \apj, 554, 778

\bibitem[{{Lima Neto} {et~al.}(1999){Lima Neto}, {Gerbal}, \&
  {M{\'a}rquez}}]{LimaNeto99}
{Lima Neto}, G.~B., {Gerbal}, D., \& {M{\'a}rquez}, I. 1999, \mnras, 309, 481

\bibitem[{{Lo Faro} {et~al.}(2017){Lo Faro}, {Buat}, {Roehlly},
  {Alvarez-Marquez}, {Burgarella}, {Silva}, \& {Efstathiou}}]{LoFaro2017}
{Lo Faro}, B., {Buat}, V., {Roehlly}, Y., {et~al.} 2017, \mnras, 472, 1372

\bibitem[{{Lynden-Bell}(1975)}]{Lynden-Bell}
{Lynden-Bell}, D. 1975, Vistas in Astronomy, 19, 299

\bibitem[{{Maraston}(2005)}]{Maraston2005}
{Maraston}, C. 2005, \mnras, 362, 799

\bibitem[{{Marigo}(2001)}]{Marigo2001}
{Marigo}, P. 2001, \aap, 370, 194

\bibitem[{{Mathews} \& {Baker}(1971)}]{Mathews+Baker}
{Mathews}, W.~G. \& {Baker}, J.~C. 1971, \apj, 170, 241

\bibitem[{{Mathis} {et~al.}(1977){Mathis}, {Rumpl}, \&
  {Nordsieck}}]{Mathis1977}
{Mathis}, J.~S., {Rumpl}, W., \& {Nordsieck}, K.~H. 1977, \apj, 217, 425

\bibitem[{{Matteucci} \& {Greggio}(1986)}]{Matteucci1986}
{Matteucci}, F. \& {Greggio}, L. 1986, \aap, 154, 279

\bibitem[{{Mocz} {et~al.}(2012){Mocz}, {Green}, {Malacari}, \&
  {Glazebrook}}]{Mocz+2012}
{Mocz}, P., {Green}, A., {Malacari}, M., \& {Glazebrook}, K. 2012, \mnras, 425,
  296

\bibitem[{{Moy}(2000)}]{thesis_Moy}
{Moy}, E. 2000, {Ph.D. thesis}, Université Paris~11

\bibitem[{{Moy} {et~al.}(2001){Moy}, {Rocca-Volmerange}, \&
  {Fioc}}]{article_Moy}
{Moy}, E., {Rocca-Volmerange}, B., \& {Fioc}, M. 2001, \aap, 365, 347

\bibitem[{{Nomoto} {et~al.}(2013){Nomoto}, {Kobayashi}, \&
  {Tominaga}}]{Nomoto+2013}
{Nomoto}, K., {Kobayashi}, C., \& {Tominaga}, N. 2013, \araa, 51, 457

\bibitem[{{Osterbrock} \& {Ferland}(2006)}]{Osterbrock+Ferland}
{Osterbrock}, D.~E. \& {Ferland}, G.~J. 2006, {Astrophysics of gaseous nebulae
  and active galactic nuclei} (University Science Books)

\bibitem[{{Paczy{\'n}ski}(1971)}]{Paczynski1971}
{Paczy{\'n}ski}, B. 1971, \actaa, 21, 417

\bibitem[{{Podigachoski} {et~al.}(2016){Podigachoski}, {Rocca-Volmerange},
  {Barthel}, {Drouart}, \& {Fioc}}]{Podigachoski2016}
{Podigachoski}, P., {Rocca-Volmerange}, B., {Barthel}, P., {Drouart}, G., \&
  {Fioc}, M. 2016, \mnras, 462, 4183

\bibitem[{{Portinari} {et~al.}(1998){Portinari}, {Chiosi}, \&
  {Bressan}}]{Portinari1998}
{Portinari}, L., {Chiosi}, C., \& {Bressan}, A. 1998, \aap, 334, 505

\bibitem[{{Pozzetti} {et~al.}(2010){Pozzetti}, {Bolzonella}, {Zucca},
  {Zamorani}, {Lilly}, {Renzini}, {Moresco}, {Mignoli}, {Cassata}, {Tasca},
  {Lamareille}, {Maier}, {Meneux}, {Halliday}, {Oesch}, {Vergani}, {Caputi},
  {Kova{\v c}}, {Cimatti}, {Cucciati}, {Iovino}, {Peng}, {Carollo}, {Contini},
  {Kneib}, {Le F{\'e}vre}, {Mainieri}, {Scodeggio}, {Bardelli}, {Bongiorno},
  {Coppa}, {de la Torre}, {de Ravel}, {Franzetti}, {Garilli}, {Kampczyk},
  {Knobel}, {Le Borgne}, {Le Brun}, {Pell{\`o}}, {Perez Montero},
  {Ricciardelli}, {Silverman}, {Tanaka}, {Tresse}, {Abbas}, {Bottini}, {Cappi},
  {Guzzo}, {Koekemoer}, {Leauthaud}, {Maccagni}, {Marinoni}, {McCracken},
  {Memeo}, {Porciani}, {Scaramella}, {Scarlata}, \& {Scoville}}]{Pozzetti+2010}
{Pozzetti}, L., {Bolzonella}, M., {Zucca}, E., {et~al.} 2010, \aap, 523, A13

\bibitem[{{Prantzos} \& {Silk}(1998)}]{Prantzos+Silk}
{Prantzos}, N. \& {Silk}, J. 1998, \apj, 507, 229

\bibitem[{{Rauch}(2003)}]{Rauch2003}
{Rauch}, T. 2003, \aap, 403, 709

\bibitem[{{Reimers}(1975)}]{Reimers1975}
{Reimers}, D. 1975, M\'emoires de la Soci\'et\'e Royale des Sciences de
  Li\`ege, 8, 369

\bibitem[{{Renzini}(1981)}]{Renzini1981}
{Renzini}, A. 1981, in Astrophysics and Space Science Library, Vol.~88,
  Physical Processes in Red Giants, ed. I.~{Iben}, Jr. \& A.~{Renzini},
  431--446

\bibitem[{{Renzini} \& {Buzzoni}(1983)}]{RenziniBuzzoni}
{Renzini}, A. \& {Buzzoni}, A. 1983, \memsai, 54, 739

\bibitem[{{Rocca-Volmerange} {et~al.}(2013){Rocca-Volmerange}, {Drouart}, {De
  Breuck}, {Vernet}, {Seymour}, {Wylezalek}, {Lehnert}, {Nesvadba}, \&
  {Fioc}}]{RoccaVolmerange2013}
{Rocca-Volmerange}, B., {Drouart}, G., {De Breuck}, C., {et~al.} 2013, \mnras,
  429, 2780

\bibitem[{{Rocca-Volmerange} \& {Fioc}(1999)}]{RVF99}
{Rocca-Volmerange}, B. \& {Fioc}, M. 1999, \apss, 269, 233

\bibitem[{{Rocca-Volmerange} {et~al.}(1981){Rocca-Volmerange}, {Lequeux}, \&
  {Maucherat-Joubert}}]{RV81}
{Rocca-Volmerange}, B., {Lequeux}, J., \& {Maucherat-Joubert}, M. 1981, \aap,
  104, 177

\bibitem[{{Sch\"onberner}(1983)}]{Schoenberner1983}
{Sch\"onberner}, D. 1983, \apj, 272, 708

\bibitem[{{Sellgren}(1984)}]{Sellgren1984}
{Sellgren}, K. 1984, \apj, 277, 623

\bibitem[{{Silva} {et~al.}(1998){Silva}, {Granato}, {Bressan}, \&
  {Danese}}]{Silva1998}
{Silva}, L., {Granato}, G.~L., {Bressan}, A., \& {Danese}, L. 1998, \apj, 509,
  103

\bibitem[{{Sommer-Larsen} {et~al.}(2003){Sommer-Larsen}, {G{\"o}tz}, \&
  {Portinari}}]{Sommer-Larsen+2003}
{Sommer-Larsen}, J., {G{\"o}tz}, M., \& {Portinari}, L. 2003, \apj, 596, 47

\bibitem[{{Spitzer}(1978)}]{Spitzer}
{Spitzer}, L. 1978, {Physical processes in the interstellar medium} (New York
  Wiley-Interscience)

\bibitem[{{Stasi{\'n}ska}(1984)}]{Stasinska1984}
{Stasi{\'n}ska}, G. 1984, \aaps, 55, 15

\bibitem[{{Steidel} {et~al.}(2016){Steidel}, {Strom}, {Pettini}, {Rudie},
  {Reddy}, \& {Trainor}}]{Steidel2016}
{Steidel}, C.~C., {Strom}, A.~L., {Pettini}, M., {et~al.} 2016, \apj, 826, 159

\bibitem[{{Tantalo} {et~al.}(1996){Tantalo}, {Chiosi}, {Bressan}, \&
  {Fagotto}}]{TCBF}
{Tantalo}, R., {Chiosi}, C., {Bressan}, A., \& {Fagotto}, F. 1996, \aap, 311,
  361

\bibitem[{{Thielemann} {et~al.}(1986){Thielemann}, {Nomoto}, \&
  {Yokoi}}]{Thielemann1986}
{Thielemann}, F.-K., {Nomoto}, K., \& {Yokoi}, K. 1986, \aap, 158, 17

\bibitem[{{Tinsley}(1972)}]{Tinsley1972}
{Tinsley}, B.~M. 1972, \aap, 20, 383

\bibitem[{{Tosi}(1988)}]{Tosi}
{Tosi}, M. 1988, \aap, 197, 33

\bibitem[{{Tsai} \& {Mathews}(1995)}]{Tsai1995}
{Tsai}, J.~C. \& {Mathews}, W.~G. 1995, \apj, 448, 84

\bibitem[{{Tumlinson} {et~al.}(2017){Tumlinson}, {Peeples}, \&
  {Werk}}]{Tumlinson+2017}
{Tumlinson}, J., {Peeples}, M.~S., \& {Werk}, J.~K. 2017, \araa, 55, 389

\bibitem[{{Twarog}(1980)}]{Twarog}
{Twarog}, B.~A. 1980, \apj, 242, 242

\bibitem[{{van den Hoek} \& {Groenewegen}(1997)}]{vdH1997}
{van den Hoek}, L.~B. \& {Groenewegen}, M.~A.~T. 1997, \aaps, 123, 305

\bibitem[{{V{\'a}rosi} \& {Dwek}(1999)}]{Varosi+Dwek}
{V{\'a}rosi}, F. \& {Dwek}, E. 1999, \apj, 523, 265

\bibitem[{{Walcher} {et~al.}(2011){Walcher}, {Groves}, {Budav{\'a}ri}, \&
  {Dale}}]{Walcher2011}
{Walcher}, J., {Groves}, B., {Budav{\'a}ri}, T., \& {Dale}, D. 2011, \apss,
  331, 1

\bibitem[{{Weingartner} \& {Draine}(2001)}]{Weingartner2001}
{Weingartner}, J.~C. \& {Draine}, B.~T. 2001, \apj, 548, 296

\bibitem[{{Westera} {et~al.}(2002){Westera}, {Lejeune}, {Buser}, {Cuisinier},
  \& {Bruzual}}]{Westera2002}
{Westera}, P., {Lejeune}, T., {Buser}, R., {Cuisinier}, F., \& {Bruzual}, G.
  2002, \aap, 381, 524

\bibitem[{{Woosley} \& {Weaver}(1995)}]{Woosley1995}
{Woosley}, S.~E. \& {Weaver}, T.~A. 1995, \apjs, 101, 181

\bibitem[{{Worthey}(1994)}]{Worthey}
{Worthey}, G. 1994, \apjs, 95, 107

\bibitem[{{Xilouris} {et~al.}(1999){Xilouris}, {Byun}, {Kylafis}, {Paleologou},
  \& {Papamastorakis}}]{Xilouris1999}
{Xilouris}, E.~M., {Byun}, Y.~I., {Kylafis}, N.~D., {Paleologou}, E.~V., \&
  {Papamastorakis}, J. 1999, \aap, 344, 868

\bibitem[{{Zubko} {et~al.}(2004){Zubko}, {Dwek}, \& {Arendt}}]{Zubko2004}
{Zubko}, V., {Dwek}, E., \& {Arendt}, R.~G. 2004, \apjs, 152, 211

\end{thebibliography}
\begin{appendix}
  \section{Adaptation of the calculations by
    \citet{Spitzer} to the case of an inner cavity%
    \label{sec:cavity}}
  Here, we consider the \HII~region created by a point-like star cluster
  in a cloud of gas and dust.
  Except for a spherical inner cavity of radius~$\rCav$ centered
  on the cluster, the cloud is assumed to be homogeneous:
  the overall number density $\nH$ of hydrogen atoms,
  whether neutral or ionized, is constant, as is the dust-to-gas ratio.
  In the dust-free case, we already know from the \textsc{Cloudy} results
  (see Sect.~\ref{sec:neb_em_clouds}) the value of the Str\"omgren radius
  $\rSz$, that is, the outer radius of the spherical shell of ionized hydrogen.
  Our goal here is to estimate the Str\"omgren radius $\rSd$
  ($\mathord{<}\,\rSz$) when dust is present, and, from that,
  the fraction $\Upsilon$ of Lyman continuum photons absorbed
  in the cloud by gas rather than by dust.
  To this purpose, we have adapted to the case where an inner cavity is present
  in the \HII~region the calculations in sect.~5.1.c of \citet{Spitzer};
  we refer in the following to the equations in this work by prefixing
  the characters ``Sp.\@'' to the equation number.

  Let us first restate the most important assumptions and simplifications
  made in \citet{Spitzer}:
  \begin{enumerate}
  \item
    The \HII~region contains only hydrogen and dust;
  \item
    The \HII~region
    is ionization-bounded, and the transition zone from the almost fully
    ionized medium to the neutral one is very thin;
  \item
    All Lyman continuum photons are considered to be at a frequency just above
    the Lyman limit;
  \item
    The standard case~B \citep[see][sect.~4.2]{Osterbrock+Ferland}
    holds for the recombination of hydrogen.
  \end{enumerate}
  Because the asymmetry parameter is close to~$1$ at short wavelengths,
  we moreover assume here that dust scattering is only forward.
  This does not make the calculations more complicated since one just has
  to replace the mass extinction coefficient of dust,
  $\kappa_\dust^\extin$, by the absorption one, $\kappa_\dust^\absor$.
  Some of these approximations and assumptions are highly debatable,
  but, given all the uncertainties on the physics of dusty \HII~regions
  (the sublimation of grains, among others), a more complicated modeling
  does not seem to be worthwhile for this paper.
  
  The probability per unit time that a given hydrogen atom at a distance~$r$
  from the cluster becomes ionized is
  \begin{equation}
    \label{eq:ion_prob}
    p(r) = \frac{\kappaH\mult \NLC\mult \neper^{-\tau(r)}}{4\mult \pi\mult r^2}\,,
    \qquad\text{(Sp.~5-24)}
  \end{equation}
  where $\kappaH$ is the ionizing cross-section of an hydrogen atom,
  $\NLC$~is the number rate of Lyman continuum photons emitted by the central
  source, and $\tau(r)$ is the absorption optical depth by
  both hydrogen and dust from the source up to the distance~$r$.
  The ionization balance is given by
  \begin{equation}
    \label{eq:balance}
    (1-\xi[r])\mult p(r) = \xi(r)\mult \nElec(r)\mult \alphaB(r)\,,
    \qquad\text{(Sp.~5-2)}
  \end{equation}
  where $\xi(r)$ is the fraction of ionized hydrogen atoms,
  $\nElec(r)$ is the number density of electrons,
  and
  $\alphaB(r)$ is the case~B recombination coefficient for hydrogen.
  Since $\nElec(r) = \xi(r)\mult \nH$ and $\xi(r) \approx 1$
  in the \HII~region, one may rewrite Eq.~\eqref{eq:balance} as
  \begin{equation}
    \label{eq:balance2}
    (1-\xi[r])\mult p(r) \approx \nH\mult \alphaB\,,
  \end{equation}
  where $\alphaB$ is now treated as a constant throughout the ionized medium.

  The absorption optical depth by hydrogen $\tauH(r)$ from the source
  up to the distance~$r$ obeys the relation
  \begin{equation}
    \label{eq:dtau}
    \df\tauH = (1-\xi[r])\mult \nH\mult \kappaH\mult \df r.
    \qquad\text{(Sp.~5-16)}
  \end{equation}
  With an inner cavity, eq.~(Sp.~5-21) becomes
  \begin{equation}
    \label{eq:NLC}
    \NLC = \frac{4\mult \pi}{3}\mult \Bigl(\rSz^3 - \rCav^3\Bigr)
    \mult \langle \nElec\mult n_{\mathrm{p}} \mult \alphaB\rangle
    \approx \frac{4\mult \pi}{3}\mult \Bigl(\rSz^3 - \rCav^3\Bigr)
    \mult \nH^2\mult \alphaB\,,
  \end{equation}
  where $\langle\cdot\rangle$ denotes a volume-average and
  $\xi$ has been approximated to~$1$ in the \HII~region.

  Combining Eqs.~\eqref{eq:ion_prob}, \eqref{eq:balance2} and \eqref{eq:NLC},
  one obtains
  \begin{equation}
    \label{eq:1-xi}
    (1-\xi[r])\mult \kappaH\mult \nH\mult \rSz
    = \frac{3\mult \neper^{\tau(r)}\mult \gamma^2(r)}{1-\gammaCav^3}\,,
  \end{equation}
  with $\gamma(r) \coloneqq r/\rSz$ and $\gammaCav \coloneqq \gamma(\rCav)$.
  Since
  \begin{equation}
    \label{eq:dtau2}
    \df\tauH = (1-\xi[r])\mult \kappaH\mult \nH\mult \rSz\mult \df\gamma
  \end{equation}
  from Eq.~\eqref{eq:dtau} and the definition of $\gamma$,
  one obtains from Eqs.~\eqref{eq:1-xi} and \eqref{eq:dtau2} that
  \begin{equation}
    \label{eq:dtau3}
    \df\tauH = \frac{3\mult \neper^{\tau(r)}\mult \gamma^2(r)}{1-\gammaCav^3}
    \mult \df\gamma.
  \end{equation}

  Let $\taud(r)$ be the absorption optical depth by dust from the source
  up to the distance~$r$.
  One has
  \begin{equation}
    \label{eq:taud}
    \taud(r) = \kappa_\dust^\absor\mult \mu_\dust\mult (r-\rCav)
    = (\gamma[r] - \gammaCav)\mult \tauSd\,,
  \end{equation}
  where $\mu_\dust$ is the mass density of dust in the \HII~region (related to
  $\nH$ by the dust-to-hydrogen ratio derived from the evolution of the galaxy),
  and
  \begin{equation}
    \tauSd \coloneqq \kappa_\dust^\absor\mult \mu_\dust\mult \rSz.
  \end{equation}
  Since $\tau(r) = \tauH(r) + \taud(r)$, Eq.~\eqref{eq:dtau3} gives
  \begin{equation}
    \label{eq:dtau4}
    \neper^{-\tauH(r)}\mult \df\tauH
    = \frac{3\mult \neper^{\taud(r)}\mult \gamma^2(r)}{1-\gammaCav^3}
    \mult \df\gamma.
  \end{equation}

  To compute $\rSd$, let us integrate
  this equation from $\rCav$ to $\rSd$, as in \citet{Spitzer}.
  The left-hand side is
  \begin{equation}
    \int_{\tauH(r)=\tauH(\rCav)}^{\tauH(\rSd)} \neper^{-\tauH(r)}\mult \df\tauH
    = \neper^{-\tauH(\rCav)} - \neper^{-\tauH(\rSd)}
    \approx 1\,,
  \end{equation}
  since $\tauH(\rCav) = 0$ and (by definition of $\rSd$) $\tauH(\rSd) \gg 1$.
  The value of $\gammaSd \coloneqq \gamma(\rSd)$ is therefore the unique
  solution in the interval $[\gammaCav, 1]$ of $f(\gammaSd) = 0$, with
  \begin{align}
    \!
    f(\gammaSd)
    &\coloneqq \frac{3\mult \neper^{-\gammaCav\mult\tauSd}}{1-\gammaCav^3}\mult
    \int_{\gamma=\gammaCav}^{\gammaSd}\neper^{\gamma\mult\tauSd}\mult \gamma^2
    \mult \df\gamma -1
    \notag\\
    &= 3\mult\frac{\neper^{\epsilonSd-\epsilonCav}
      \mult \bigl(\epsilonSd^2-2\mult \epsilonSd+2\bigr)
      - \bigl(\epsilonCav^2-2\mult \epsilonCav+2\bigr)}{
      \bigl(1-\gammaCav^3\bigr)\mult \tauSd^3} - 1\,,
  \end{align}
  where $\epsilonSd \coloneqq \gammaSd\mult \tauSd$ and
  $\epsilonCav \coloneqq \gammaCav\mult \tauSd$.
  For $\tauSd \ll 1$, the evaluation of $f(\gammaSd)$ is numerically unstable.
  It is then safer to use instead its first-order expansion in $\tauSd$
  near~$0$,
  \begin{equation}
    f_1(\gammaSd) = \frac{1}{1-\gammaCav^3}\mult \Biggl(\gammaSd^3-1 +
    \frac{\tauSd/4}{\gammaCav^4-4\mult \gammaCav\mult \gammaSd^3
      +3\mult \gammaSd^4}\Biggr).
  \end{equation}

  Finally, the fraction of Lyman continuum photons absorbed in the \HII~region
  by hydrogen atoms and not grains is the volume reduction factor
  caused by dust,
  \begin{equation}
    \Upsilon = \frac{4\mult \pi\mult \Bigl(\rSd^3-\rCav^3\Bigr)/3}{
      4\mult \pi\mult \Bigl(\rSz^3-\rCav^3\Bigr)/3}
    = \frac{\gammaSd^3-\gammaCav^3}{1-\gammaCav^3}.
  \end{equation}
\end{appendix}
\end{document}